\documentclass[11pt]{article}
\usepackage{amsmath,amssymb,amsfonts,amsthm,authblk}
\usepackage{braket}
\usepackage{moreverb,rotating,graphics,bm}
\usepackage[T1]{fontenc}
\usepackage{epsfig}
\usepackage{dsfont}
\usepackage{subcaption}
\usepackage{ulem}
\usepackage{cancel}
\usepackage{xcolor}
\usepackage[sorting=none  , 
style=phys,
maxbibnames=99,
style=chem-acs,
autocite=superscript]{biblatex} 
\usepackage{mathtools}
\usepackage{graphicx}

\allowdisplaybreaks

\newtheorem{lemma}{Lemma}

\def\tr{{\rm Tr \,}}

\def\R{{\mathbb R}}
\def\C{{\mathbb C}}

\def\1{{\mathds{1}}}

\newcommand \dps{\displaystyle }

\newcommand{\tabincell}[2]{\begin{tabular}{@{}#1@{}}#2\end{tabular}}

\graphicspath{{figures/}}

\title{Critical point search and linear response theory for computing electronic excitation energies of molecular systems. \\ $\mbox{Part I: General framework, application to Hartree-Fock and DFT}$}

\author{Laura Grazioli}
\author{Yukuan Hu} 
\author{Eric Canc\`es}

\affil{CERMICS, Ecole des Ponts - Institut Polytechnique de Paris and Inria, \qquad\qquad 6-8 avenue Blaise Pascal, Cit\'e Descartes, 77455 Marne-la-Vall\'ee, France}

\bibliography{ref}
\begin{document}

\maketitle

\begin{abstract}
Computing excited states of many-body quantum Hamiltonians is a fundamental challenge in computational physics and chemistry, with state-of-the-art methods broadly classified into variational (critical point search) and linear response approaches. The Kähler manifold formalism provides a uniform framework which naturally accommodates both strategies for a wide range of variational models, including Hartree-Fock, CASSCF, Full CI, and adiabatic TDDFT. In particular, this formalism leads to a systematic and straightforward way to obtain the final equations of linear response theory for nonlinear models, which provides, in the case of mean-field models (Hartree-Fock and DFT), a simple alternative to Casida's derivation. We detail the mathematical structure of Hamiltonian dynamics on Kähler manifolds, establish connections to standard quantum chemistry equations, and provide theoretical and numerical comparisons of excitation energy computation schemes at the Hartree-Fock level.
\end{abstract}

\section{Introduction}

Computing excited states of many-body quantum Hamiltonians is one of the biggest challenges in computational physics and chemistry.\supercite{Gonzlez2011}

\medskip

The main approaches for computing approximate excited states can be classified into two categories:
\begin{enumerate}
    \item variational or bivariational methods: computation of critical points of energy functionals on differentiable manifolds or algebraic varieties. This category encompasses state-specific and state-averaged methods;\supercite{Docken1972,Nicolaides1984,Hirao1993, Kossoski2021,Kossoski2023,Marie2023,Saade2024}
    \item methods based on linear response theory. Linear-response Hartree-Fock\supercite{Jrgensen1970, Olsen1988,Dalgaard1982,Matsuo2001,Ebata2010,Miura2010,vanMeer2017}(LR-HF), Complete Active Space Self-Consistent Field\supercite{Berger2000,Stlring2002,HelmichParis2019benchmarks, HelmichParis2019} (LR-CASSCF), time-dependent density functional theory\supercite{Runge1984,Gross1985, Onida2002,CASIDA1995,Casida1998,Casida2005} (LR-TDDFT), and equation-of-motion coupled cluster\supercite{Mukherjee1979,Sekino1984,Nooijen1995,Nooijen1997,Nooijen1997anewmethod,Geertsen1989,Stanton1993,koch1990coupled,Comeau1993} (EOM-CC) fall into this category.
\end{enumerate}
All these approaches coincide for the reference electronic Schr\"odinger equation set on the continuous position space $\bigwedge^N L^2(\R^3;\C^2)$, as well as for the Full Configuration Interaction (FCI) approximation,\supercite{Knowles1984,Olsen1990} but differ when further approximations are made. This contribution aims at analyzing and comparing these methods from a theoretical viewpoint.

\medskip

For this purpose, we will use the K\"ahler manifold formalism, allowing one to conveniently formulate both critical point search (CP) and linear response theory (LR)  for any variational model,\supercite{Rowe1980} including any variant of HF \supercite{Hartree1928,Slater1928, fock1930naherungsmethode,hartree1935self} (restricted, unrestricted, general, restricted open-shell), CASSCF,\supercite{Roos1980} and FCI\supercite{Knowles1984,Olsen1990} methods, in a unified and compact framework. It also applies to adiabatic TDDFT,\supercite{marques2006time,marques2012fundamentals} as this method has an intrinsic Hamiltonian K\"ahler structure, although it is not a variational approximation of the reference electronic Schr\"odinger equation. Although the K\"ahler manifold formalism was previously used in the physics literature to study infinite-dimensional Hamiltonian systems, including the time-dependent Schr\"odinger equation,\supercite{ChernoffMarsden1974} or to compute excitation modes of mean-field models (normal mode analysis),\supercite{Rowe1980} it is very little used in quantum chemistry. The only quantum chemistry articles we are aware of following a similar approach are the works by Arponen\supercite{Arponen1983,Arponen1991} on Coupled-Cluster theory, and the recent work by Kvaal and coworkers\supercite{Kvaal2025} aiming at better understanding the structure of approximation methods for time-dependent
quantum dynamics.

\medskip

We will concentrate on the FCI and HF theories in Part I of this contribution. The cases of more advanced variational approximations such as CASSCF, DMRG, and bivariational approximations such as CC,\supercite{Cizek1980,Cizek1991,Bartlett2007} will be dealt with in Parts II and III.

\medskip

\begin{figure}[h]
    \centering
    \includegraphics[width=0.5\linewidth]{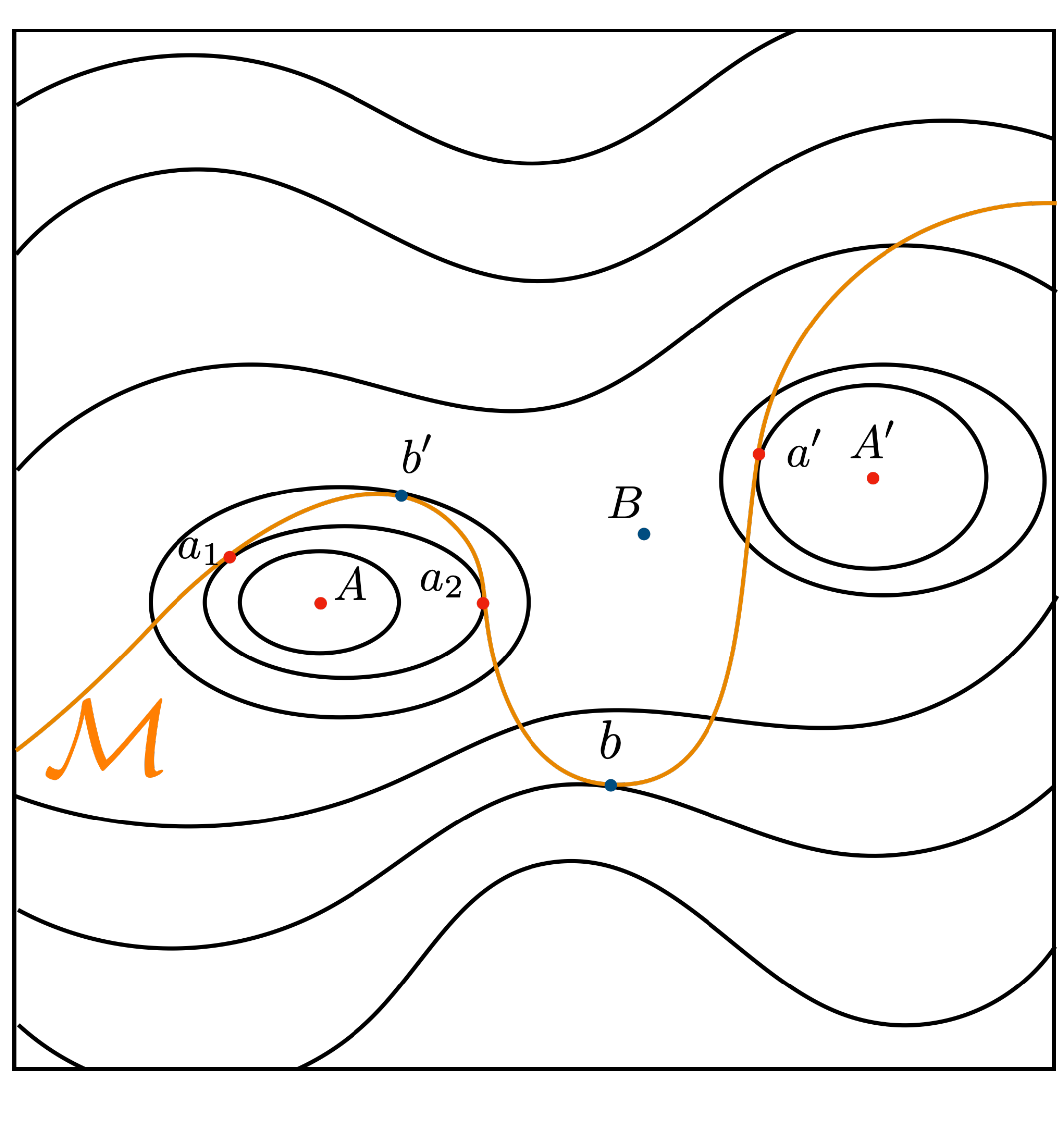}
    \caption{Contour of a 2D function $E:(0,1) \times (0,1) \to \R$ with one global minimizer ($A$), one local, non-global minimizer ($A'$) and one saddle point with Morse index 1 ($B$). The restriction $\mathcal E$ of this function to the one-dimensional manifold $\mathcal M$ (indicated by the orange curve) has 3 global minimizers ($a_1$, $a_2$, $a'$), 0 local, non-global, minimizers, and 2 critical points ($b$, $b^\prime$) with a Morse index 1, which are also local maxima since $\mathcal M$ is one-dimensional. The points $a_1$ and $a_2$ can be considered as reasonable approximations of $A$, $a'$ as a reasonable approximation of $A'$, $b$ as a reasonable approximation of $B$. The point $b^\prime$ is a spurious critical point, which does not approximate well any saddle point of~$E$.}
    \label{fig:cartoon}
\end{figure}

If $\mathcal M$ is a K\"ahler manifold (the definition and properties of this mathematical structure will be recalled in Section~\ref{sec:Kahler}), and $\mathcal E : \mathcal M \to \R$ an energy functional over $\mathcal M$, the Hamiltonian K\"ahler dynamics is defined as
\begin{equation}\label{eq:dynamics_0}
{\frac{dx}{dt}(t) = J_{x(t)} {\rm grad}_{\mathcal M}\mathcal E(x(t))},
\end{equation} 
where ${\rm grad}_{\mathcal M}\mathcal E(x)$ is the Riemannian gradient of $\mathcal E$ at point $x \in \mathcal M$, and  $J_x$ the so-called symplectic operator at point $x$. 
The critical points (steady states) therefore satisfy
\begin{equation} \label{eq:critical_points_0}
{{\rm grad}_{\mathcal M}\mathcal E(x_\star)=0}.
\end{equation}
The CP approach for computing excitation energies consists in:
\begin{enumerate}
    \item finding critical points $x_0, x_1, x_2,\ldots$ of $\mathcal E$ on $\mathcal M$ that can be identified as approximate ground and excited states, labelled in such a way that $\mathcal E(x_0) \le \mathcal E(x_1) \le \mathcal E(x_2)\le \cdots$;
    \item computing the $k$-th excitation energy as the difference 
    \begin{equation} \label{eq:omega_CP}
        \omega^{\rm CP}_k := \mathcal E(x_k) - \mathcal E(x_0),
    \end{equation}
    between the energies of the approximation $x_k$ of the $k$-th excited state and the approximation $x_0$ of the ground state, i.e., the (presumably) global minimum of $\mathcal E$ on $\mathcal M$.
\end{enumerate}
Step 1 is quite involved. Indeed, in contrast with linear models such as FCI, not all critical points of nonlinear models such as HF or CASSCF are suitable approximations of ground or excited states of the reference electronic Schr\"odinger equation. In general, the energy functional $\mathcal E$ can have many spurious critical points on $\mathcal M$. Worse still, there can be mismatch between the $k$-th excited states and the critical points with a Morse index $k$, i.e., critical points where the Riemannian Hessian has exactly $k$ negative eigenvalues.\supercite{marie2023excited} This phenomenon is illustrated on a 2D toy model in Fig.~\ref{fig:cartoon}, as well as on an H$_4$ molecule in Section \ref{subsec:CP multiple solutions}. In Ref.~\cite{lewin2004solutions}, Lewin introduced a nonlinear min-max characterization of physically-relevant critical points for CASSCF with no internal orbitals. In Ref.~\cite{cances2006computing}, an algorithm for computing CASSCF approximations of first excited states based on this characterization was proposed and tested on an H$_2$ molecule. We will elaborate further on this characterization in Part II.

In LR, the excitation energies are identified with the resonant energies of the linearized dynamics around the global minimum $x_0$ of $\mathcal E$ on $\mathcal M$. This dynamics is given by
\begin{equation}  \label{eq:linear_response_11}
    {\frac{dy}{dt}(t) = J_{x_0} {\rm Hess}_{\mathcal M} \mathcal E(x_0) y(t)},
\end{equation}
where ${\rm Hess}_{\mathcal M} \mathcal E(x_0)$ is the Riemannian Hessian of $\mathcal E$ at $x_0$. Under a nondegeneracy assumption, it follows that the approximate excitation energies $\omega^{\rm LR}_k$ computed by LR are in fact the symplectic eigenvalues of ${\rm Hess}_{\mathcal M} \mathcal E(x_0)$:
\begin{equation}\label{eq:linear_response_1}
    \{\omega_k^{\rm LR}\}_{1 \le k \le \dim_{\C}(\mathcal M)} := \{ \mbox{symplectic eigenvalues of }  {\rm Hess}_{\mathcal M} \mathcal E(x_0) \}.
\end{equation}
This approach provides a systematic and conceptually simple way to derive linear-response approximations to excitation energies for any variational model with K\"ahler structure. To our knowledge, this approach was first used by Rowe and coworkers\supercite{Rowe1980} in the context of mean-field models.

\medskip

In Section~\ref{sec:Kahler}, we present the abstract framework of Hamiltonian dynamics on a generic K\"ahler manifold, and define precisely the mathematical objects arising in the fundamental Eqs. \eqref{eq:dynamics_0}-\eqref{eq:linear_response_1}. In Section~\ref{sec:Grassmann}, we consider the specific case of Hamiltonian dynamics on Grassmann manifolds
\begin{equation} \label{eq:Grassmann_1}
\mathcal M= {\rm Gr}_\C(r,n) := \left\{ \gamma \in  \C^{n \times n}_{\rm herm} \; | \; \gamma^2=\gamma, \; \tr(\gamma)=r \right\},
\end{equation}
allowing one to deal with FCI (or more generally any linear Schr\"odinger or Liouville equation on a finite-dimensional state-space), HF, and adiabatic TDDFT. In Section~\ref{subsec:geo_Grassmann}, we describe the K\"ahler structure of Grassmann manifolds. In Section~\ref{subsec:dyn_grass}, we give more explicit formulations of Eqs.~\eqref{eq:dynamics_0}-\eqref{eq:linear_response_1} in the special case when $\mathcal M$ is a Grassmann manifold and apply this formalism to a molecular system with $N_{\rm b}$ atomic spin-orbitals and $N$ electrons. In Section~\ref{subsec:linear_functionals}, we deal with the FCI framework, where $n=\binom{N_{\rm b}}{N}$, $r=1$, $\Gamma\in {\rm Gr}_\C(1,n) \subset \C_{\rm herm}^{n\times n}$ represents the $N$-body density matrix, and the energy functional ${\mathcal E}$ is linear, i.e., ${\mathcal E}(\Gamma)=\tr(H\Gamma)$  with $H\in\C_{\rm herm}^{n\times n}$ the Hamiltonian matrix (specified later). We show that Eq.~\eqref{eq:dynamics_0} is then equivalent to the time-dependent Schr\"odinger equation:
$$
\widehat\Gamma(t) = |\Psi(t)\rangle \langle \Psi(t)| \quad \mbox{with} \quad i \frac{d\Psi}{dt}(t) = \widehat H \Psi(t), \quad \|\Psi(t)\|=1,
$$
and that Eq.~\eqref{eq:critical_points_0} is equivalent to the time-independent Schr\"odinger equation:
$$
 \widehat\Gamma = |\Psi\rangle \langle \Psi| \quad \mbox{with} \quad \widehat H\Psi=E\Psi, \quad  \|\Psi\|=1.
$$
We also show that, denoting by $E_0$ the ground-state energy of $\widehat H$, by $\Psi_\star$ a normalized ground state, and by $\widehat\Gamma_\star=\ket{\Psi_\star}\bra{\Psi_\star}$, the symplectic eigenvalues of ${\rm Hess}_{\mathcal M}\mathcal E(\Gamma_\star)$ coincide with the eigenvalues of 
$
(\widehat H-E_0)|_{\Psi_\star^\perp}.
$
It follows that, as expected,
$$
\omega_k^{\rm CP-FCI} = \omega_k^{\rm LR-FCI}=E_k-E_0,
$$
where $E_k$ is the energy of the $k$-th FCI excited state. In Section~\ref{subsec:mean-field_models}, we consider the case of mean-field models, including HF and adiabatic TDDFT. In this case, $n=N_{\rm b}$ is the size of the basis set, $r=N$, $\gamma\in\C_{\rm herm}^{N_{\rm b}\times N_{\rm b}}$ represents the one-body reduced density matrix (1-RDM), and the energy functional ${\mathcal E}$ is nonlinear. Eq.~\eqref{eq:dynamics_0} is then equivalent to the time-dependent HF (TDHF) or adiabatic TDDFT equations, Eq.~\eqref{eq:critical_points_0} to the usual HF or Kohn-Sham equations, and Eq.~\eqref{eq:linear_response_1} leads to a straightforward derivation of the final equation providing single excitation energies. Recall that this equation is usually inferred from Casida's equations of LR-TDHF or LR-TDDFT.\supercite{CASIDA1995} In Section~\ref{sec:weakly_interacting}, we make use of the compact formalisms \eqref{eq:dynamics_0}-\eqref{eq:linear_response_1} to provide a theoretical comparison of the electronic excitation energy computation methods based on CP on the one-hand and LR on the other hand, at the unrestricted Hartree-Fock (UHF) \supercite{Pratt1956,Marshall1961,Amos1964,Fukutome1971} level of theory in the weakly interacting regime. More precisely, we consider a family of energy functionals of the form
$$
\mathcal{E}^{\rm UHF}_\eta(\gamma_\alpha,\gamma_\beta):={\rm Tr}(h\gamma_\alpha) + {\rm Tr}(h\gamma_\beta)+\eta {\mathcal E}_{\rm Hx}^{\rm UHF}(\gamma_\alpha,\gamma_\beta),
$$
on the manifold $\mathcal M={\rm Gr}_\C(N_\alpha,N_{\rm b}) \times {\rm Gr}_\C(N_\beta,N_{\rm b})$, where $N_\alpha$ and $N_\beta$ are the numbers of spin-up and spin-down electrons respectively, $h \in \C^{N_{\rm b} \times N_{\rm b}}_{\rm herm}$ the one-electron Hamiltonian matrix, ${\mathcal E}^{\rm UHF}_{\rm Hx}$ the Hartree plus exchange UHF energy functional and $\eta$ the coupling parameter. 
We compare the UHF electronic excitation energies $\omega^{\rm UHF,CP}_k(\eta)$ and  $\omega^{\rm UHF,LR}_k(\eta)$ obtained by CP and LR respectively for small values of $\eta$. In Section~\ref{sec:numerics}, we report numerical results supporting our theoretical analysis.

\section{Hamiltonian dynamics on K\"ahler manifolds}
\label{sec:Kahler}

\noindent
In this section, we intend to provide a brief introduction to K\"ahler manifolds and associated Hamiltonian dynamics. We refer interested readers to, e.g., the monographs~\cite{huybrechts2005complex,Moroianu_2007,wells2008differential}.

\par A K\"ahler manifold is a complex manifold $\mathcal M$ of complex dimension $n={\rm dim}_\C(\mathcal M)$, endowed with a positive definite Hermitian form $\langle \bullet,\bullet \rangle_\bullet$, which allows one to endow $\mathcal M_\R$ (the manifold $\mathcal M$ seen this time as a real $2n$-dimensional manifold) with
\begin{enumerate}
\item a Riemannian structure with metric defined by
$$
\forall u,v \in T_x\mathcal M_\R, \quad g_x(u,v) := \mbox{Re}(\langle u,v \rangle_x);
$$
\item a symplectic structure with symplectic form defined by 
$$
\forall u,v \in T_x\mathcal M_\R, \quad \omega_x(u,v) := \mbox{Im}(\langle u,v \rangle_x).
$$
\end{enumerate}
The symplectic operator $J_x:T_x\mathcal M_\R \to T_x \mathcal M_\R$ is defined by
$$
\forall u,v \in T_x\mathcal M_\R, \quad g_x(u,v)= \omega_x(u,J_xv).
$$ 
It is easy to check that, 
\begin{itemize}
    \item the bilinear forms $g_x$ and $\omega_x$ are respectively real-symmetric and real-antisymmetric on $T_x\mathcal M_\R$;
    \item $J_x$ is invertible, squares to $-{\rm Id}_{T_x\mathcal M_{\R}}$, and transforms any vector $u \in T_x\mathcal M_\R$ into a vector $J_xu \in T_x\mathcal M_\R$ orthogonal to $u$:
    \begin{equation}\label{eq:Jxu_ortho_u}
    g_x(u,J_x u)= \omega_x(u,J_x^2u) = \omega_x(u,-u) = - \omega_x(u,u)=0;
    \end{equation}
    \item for any $u$, $v\in T_x\mathcal{M}_{\R}$, 
\begin{equation}
    \label{eqn:properties of Kahler manifold}
    g_x(J_x u,J_x v)=g_x(u,v),\quad \omega_x(J_x u,J_x v)=\omega_x(u,v),\quad \omega_x(u,v)=-g_x(u,J_x v).
\end{equation}
\end{itemize}

\par Let $\mathcal E:\mathcal M_\R \to \R$ be a smooth real-valued function on $\mathcal M_\R$. We denote respectively by 
$$
d\mathcal E: \mathcal M_\R \to (T\mathcal M_\R)^* \qquad \mbox{and} \qquad {\rm grad}_{\mathcal M} \mathcal E: \mathcal M_\R \to T\mathcal M_\R
$$
the differential of $\mathcal E$ and its Riemannian gradient. For each $x \in \mathcal M_\R$, $d\mathcal E(x)$ is the unique $\R$-linear form on $T_x\mathcal M_\R$ such that for any smooth path $c:[-1,1] \to \mathcal M_\R$ drawn on $\mathcal M_\R$ satisfying $c(0)=x$ and $\dot c(0)=v \in T_x\mathcal M_\R$,
$$
\mathcal E(c(t))=\mathcal E(x) + t \, d\mathcal E(x)(v)+o(t),
$$
and ${\rm grad}_{\mathcal M}E(x)$ the unique vector of $T_x\mathcal M_\R$ such that
$$
\forall v \in T_x\mathcal M_\R, \quad d\mathcal E(x)(v) = g_x\left({\rm grad}_{\mathcal M}\mathcal E(x),v\right).
$$

\par
\begin{figure}[h]
    \centering
    \includegraphics[width=0.5\linewidth]{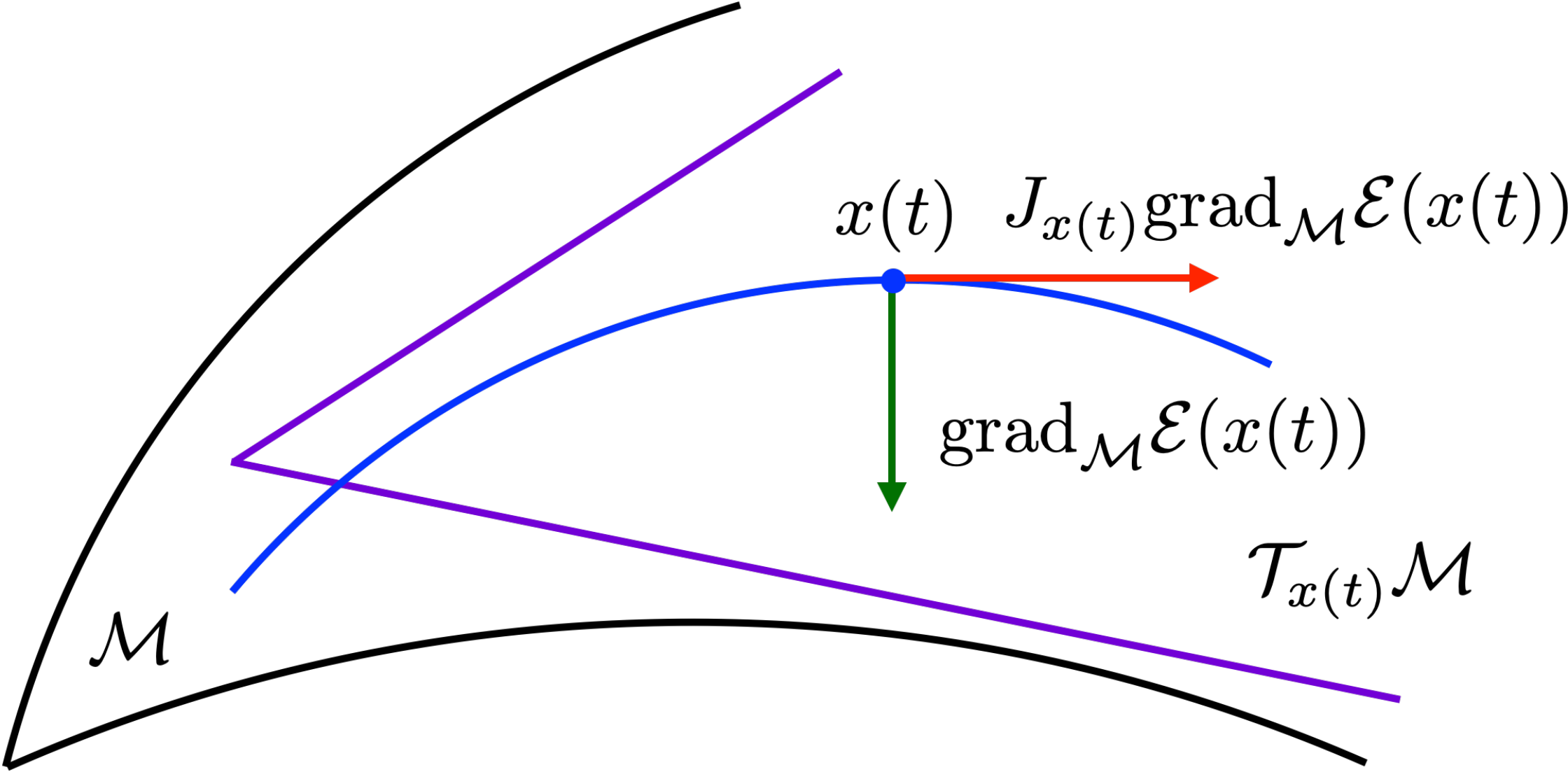}
    \caption{Geometric interpretation of a trajectory $t \mapsto x(t)$ of the Hamiltonian dynamics \eqref{eq:dynamics} on a K\"ahler manifold $\mathcal M$.}
    \label{fig:manifold}
\end{figure}

 The Hamiltonian dynamics on the symplectic manifold $(\mathcal M_\R,J_\bullet)$ is defined by
\begin{equation}\label{eq:dynamics}
{\frac{dx}{dt}(t) = J_{x(t)} {\rm grad}_{\mathcal M}\mathcal E(x(t))}.
\end{equation} 
A geometric interpretation of the above dynamics is sketched in Fig.~\ref{fig:manifold}. It follows from Eq.~\eqref{eq:Jxu_ortho_u} that at each time $t$, the velocity vector $\frac{dx}{dt}(t)$ is orthogonal to the Riemannian gradient of the energy functional $\mathcal E$ at point $x(t)$. As a consequence, the energy $\mathcal E$ is a conserved quantity.

Before proceeding further, let us point out that in the simple case of the flat space $\mathcal M=\C^n$ endowed with its natural complex Hermitian form $\langle \bm z, \bm z'\rangle = \bm z^* \bm z'$, the metric, symplectic form, and symplectic operator are respectively given by 
$$
g_{\bm z_0}(\bm z,\bm z') = {\rm Re}(\bm z^* \bm z'), \qquad 
\omega_{\bm z_0}(\bm z,\bm z') = {\rm Im}(\bm z^* \bm z'),\qquad J_{\bm z_0} \bm z=i\bm z.
$$
Decomposing any vector $\bm z \in \C^n$ into its real and imaginary parts as  
$$
\bm z=\bm q + i \bm p \quad \mbox{with} \quad \bm q,\bm p \in \R^n,
$$
and introducing the classical Hamiltonian $H_{\rm cl}$ defined on the phase space $\R^n \times \R^n$ by 
$$
H_{\rm cl}(\bm q,\bm p)=\mathcal E(\bm q+i \bm p),
 $$
 the dynamics
 $$
\frac{d\bm z}{dt}(t) = J_{\bm z(t)} {\rm grad}_{\mathcal M}\mathcal E(\bm z(t))
$$
is equivalent to the classical Hamiltonian dynamics
\begin{equation}\label{eq:classical_HD}
\left\{ \begin{array}{l} \dps \frac{d\bm q}{dt}(t) = \frac{\partial H_{\rm cl}}{\partial \bm p}(\bm q(t),\bm p(t)), \\ \; \\
 \dps \frac{d\bm p}{dt}(t) = - \frac{\partial H_{\rm cl}}{\partial \bm q}(\bm q(t),\bm p(t)). \end{array} \right..
\end{equation}
The dynamics~\eqref{eq:dynamics} can therefore be interpreted as a generalization of the classical Hamiltonian dynamics \eqref{eq:classical_HD} on the finite-dimensional phase space $\R^n \times \R^n$ to finite or infinite-dimensional K\"ahler manifolds.

\medskip

Let us come back to the general case of a generic K\"ahler manifold.
As the operator $J_x$ is invertible, the equilibrium points (steady states) of the dynamics \eqref{eq:dynamics} are characterized by the equation
\begin{equation} \label{eq:critical_points}
{{\rm grad}_{\mathcal M}\mathcal E(x_\star)=0}.
\end{equation}
The equilibrium points are therefore the critical points of the energy functional $\mathcal E$ on the state manifold.

\medskip

The LR around an equilibrium point $x_\star$ consists in studying the linearized dynamics around $x_\star$ given by
\begin{equation}\label{eq:linear_response}
{\frac{dy}{dt}(t) = J_{x_\star} {\rm Hess}_{\mathcal M} \mathcal E(x_\star) y(t)},
\end{equation}
where ${\rm Hess}_{\mathcal M}\mathcal E(x_\star) : T_{x_\star} \mathcal M_\R \to T_{x_\star}\mathcal M_\R$ is the Riemannian Hessian of $\mathcal E$ at $x_\star$. Recall that ${\rm Hess}_{\mathcal M}\mathcal E(x_\star)$ is the unique symmetric $\R$-linear operator on  $T_{x_\star} \mathcal M_\R$ such that for all smooth paths $c:[-1,1] \to \mathcal M_\R$ satisfying $c(0)=x_\star$ and $\dot c(0)=v \in T_{x_\star}\mathcal M_\R$,
$$
\mathcal E(c(t)) = \mathcal E(x_\star) + \frac{t^2}2 \,  g_{x_\star}(v,{\rm Hess}_{\mathcal M}\mathcal E(x_\star)v) + o(t^2),
$$
and that symmetric means
$$
\forall u,v \in T_{x_\star}\mathcal M_\R, \quad g_{x_\star}(u,{\rm Hess}_{\mathcal M}\mathcal E(x_\star)v)=g_{x_\star}({\rm Hess}_{\mathcal M}\mathcal E(x_\star)u,v).
$$
Eq.~\eqref{eq:linear_response} is a Hamiltonian dynamics on the flat symplectic space $(T_{x_\star} \mathcal M_\R,J_{x_\star})$ with a quadratic Hamiltonian 
$$
\mathfrak h_{x_\star}(y):= \frac 12 g_{x_\star}(y,{\rm Hess}_{\mathcal M}\mathcal E(x_\star) y).
$$

\medskip

\par
The next step consists in introducing a system of coordinates allowing one to identify $(T_{x_\star} \mathcal M_\R,J_{x_\star})$ with the usual symplectic space $\big( \R^n \times \R^n, J := \begin{psmallmatrix} 0 & I_n \\ -I_n & 0 \end{psmallmatrix}\big)$. For this purpose, we first consider an orthonormal basis $(u_1,\ldots, u_n)$  of $T_{x_\star}\mathcal M$ for $\langle \bullet,\bullet \rangle_{x_\star}$, i.e., a family of $n$ vectors of $T_{x_\star}\mathcal M$ such that
$$
\langle u_j,u_{j'} \rangle_{x_\star} = \delta_{jj'},
$$
and $\mathcal X:={\rm Span}_\R(u_1,\ldots,u_n)$ the $n$-dimensional real vector space spanned by the $u_j$'s. We have
$$
T_{x_\star}\mathcal M_\R = \mathcal X \oplus_\R (J_{x_\star}\mathcal X),
$$
and we infer from Eq.~\eqref{eqn:properties of Kahler manifold} that in the basis $(u_1,\ldots,u_n,{-J_{x_\star}u_1,\ldots,-J_{x_\star}u_n})$ of $T_{x_\star}\mathcal M_\R$, $g_{x_\star}$ is represented by the identity matrix $I_{2n}$, $J_{x_\star}$ by the skew symmetric matrix
$$J:=\begin{pmatrix}
    0 & I_n \\
    -I_n & 0
\end{pmatrix}.
$$
The Hessian matrix ${\rm Hess}_{\mathcal M}\mathcal E(x_\star)$ is represented by the real symmetric matrix
\begin{equation} \label{eq:matrix_H*}
    \mathfrak H_\star^{\rm LR} := \left( \begin{array}{cc} H^{\rm rr} & H^{\rm ri} \\ H^{\rm ir} &   H^{\rm ii} \end{array} \right) \in \R^{2n \times 2n}_{\rm sym}
\end{equation}
with entries
\begin{align}
[H^{\rm rr}]_{jj'} &:= g_{x_\star}(u_j,{\rm Hess}_{\mathcal M}\mathcal E(x_\star) u_{j'}),  \label{eq:Hrr} \\
[H^{\rm ri}]_{jj'} &:= g_{x_\star}(u_j,{\rm Hess}_{\mathcal M}\mathcal E(x_\star) (-J_{x_\star} u_{j'})), \\
[H^{\rm ir}]_{jj'} &:= g_{x_\star}(-J_{x_\star}u_j,{\rm Hess}_{\mathcal M}\mathcal E(x_\star) u_{j'}), \\
[H^{\rm ii}]_{jj'} &:= g_{x_\star}(-J_{x_\star}u_j,{\rm Hess}_{\mathcal M} \mathcal E(x_\star) (-J_{x_\star}u_{j'})). \label{eq:Hii}
\end{align}
If $x_\star$ is a nondegenerate local minimizer of $\mathcal E$ on $\mathcal M_\R$, then the matrix $\mathfrak H_\star^{\rm LR}$ is real symmetric positive definite (SPD). According to Williamson's theorem, \supercite{williamson1936algebraic} there exists a symplectic matrix $S \in \R^{2n \times 2n}$ (i.e., a matrix in $\R^{2n \times 2n}$ such that $S^T J S = J$) such that
\begin{equation}\label{eq:frequencies}
S^T \mathfrak H_\star^{\rm LR} S = \left( \begin{array}{cc} \Omega & 0 \\ 0 & \Omega \end{array} \right) \quad \mbox{with} \quad \Omega:=\mbox{diag}(\omega_1^{\rm LR},\ldots,\omega_n^{\rm LR}), \quad  0 < \omega_1^{\rm LR} \le  \cdots \le \omega_n^{\rm LR}.
\end{equation}
The $\omega_j^{\rm LR}$'s are called the symplectic eigenvalues of $\mathfrak H_\star^{\rm LR}$. It is easy to show that these quantities are independent of the chosen orthonormal basis $(u_1,\ldots, u_n)$ of $T_{x_\star}\mathcal M$. They are therefore intrinsic and are called the symplectic eigenvalues of ${\rm Hess}_{\mathcal M}\mathcal E(x_\star)$.

\medskip

Denoting by $Y(t) \in \R^{2n}$ the column vector containing the coefficients of $y(t)$ in the basis $(u_1,\ldots,u_n,-J_{x_\star} u_1,\ldots,-J_{x_\star} u_n)$ and by 
$$
Z(t) := \left( \begin{array}{cc} \Omega^{-1/2} & 0 \\ 0 & \Omega^{1/2} \end{array} \right) S^{-1}Y(t) =: \left( \begin{array}{c} q(t) \\ p(t) \end{array} \right) \quad \mbox{with} \quad q(t),p(t) \in \R^{n},
$$
we have
$$
\frac{dZ}{dt}(t) = J \nabla \mathfrak H^{\rm HO}_\star(Z(t)) \quad \mbox{with} \quad \mathfrak H^{\rm HO}_\star(Z)=\mathfrak H^{\rm HO}_\star\left( \begin{array}{c} q \\ p \end{array} \right) =\frac 12 |p|^2 + \frac 12 q^T \Omega^2 q.
$$
The angular frequencies $\omega_1^{\rm LR},\ldots,\omega_n^{\rm LR}$ therefore represent the excitation frequencies of the system around the local minimizer $x_\star$ in the linear response regime.

\medskip

\par
Let us finally consider the special case of separable Hamiltonians of the form
\begin{equation}\label{eq:separable_Hamiltonians}
\mathfrak H_\star^{\rm LR} = \left( \begin{array}{cc} H^{\rm rr} & 0 \\ 0 &   H^{\rm ii} \end{array} \right) \in \R^{2n \times 2n}_{\rm sym}.
\end{equation}
Introducing the symplectic matrix
$$
S = \left( \begin{array}{cc} R & 0 \\ 0 & R^{-T} \end{array} \right) \quad \mbox{with} \quad R = (H^{\rm rr})^{-1/2} \left( (H^{\rm rr})^{1/2} H^{\rm ii} (H^{\rm rr})^{1/2} \right)^{1/4},
$$
we get 
\begin{equation}\label{eq:angular_frequencies_separable_Hamiltonians}
S^T \mathfrak H_\star^{\rm LR} S =  \left( \begin{array}{cc} \widetilde \Omega & 0 \\ 0 & \widetilde \Omega \end{array} \right), \quad \mbox{with} \quad \widetilde \Omega =  \left( (H^{\rm rr})^{1/2} H^{\rm ii} (H^{\rm rr})^{1/2} \right)^{1/2}.
\end{equation}
Diagonalizing $\widetilde \Omega$ readily provides the symplectic eigenvalues of $\mathfrak H_\star^{\rm LR}$.

\medskip

The K\"ahler formalism therefore provides a systematic way to derive linear response excitation energies for any variational model, which can be summarized in the right column of Table \ref{tab:comparison between CISD and LR} below. The key step is the identification of the canonical basis of the tangent space $T_{x_0}\mathcal M$ at the ground state $x_0$. The rest is just straightforward computation of matrix elements using Slater-Condon rules, and application of generic results of linear algebra.  The situation is somewhat similar to the computation of truncated CI excited states, where the key step is the choice of the canonical basis of the truncated CI space, composed of the HF ground state $\Phi_0$, the single excitations $\Phi_i^a$, the double excitations $\Phi_{ij}^{ab}$, etc.

\begin{table}[htb]
    \centering
    \caption{Steps for computing excitation energies in truncated CI (here CISD as a matter of illustration) and LR for variational models.}
    \label{tab:comparison between CISD and LR}
    \resizebox{\linewidth}{!}{\begin{tabular}{c||c}
        \bf CISD & \bf LR for variational models \\[2pt]\hline\hline
        \tabincell{c}{Compute an HF ground state\\$\dps\Phi_0\in\mathop{\rm argmin}_{\Phi} E^{\rm HF}(\Phi)$} & \tabincell{c}{Compute a ground state of the model\\$\dps x_0 \in \mathop{\rm argmin}_{x \in \mathcal M} \mathcal E(x)$} \\[15pt]\hline\hline
        \tabincell{c}{Identity the CISD subspace canonical basis\\$\mathcal H^{\rm CISD}= \mbox{Span}(\Phi_0,\Phi_i^a, \Phi_{ij}^{ab})$} & \tabincell{c}{Identify the canonical basis of $T_{x_0}\mathcal{M}_{\R}$\\$T_{x_0}\mathcal M_\R= \mbox{Span}_\R(u_1,\ldots,u_n,-J_{x_0}u_1,\ldots,-J_{x_0}u_n)$}\\[15pt]\hline\hline
        \tabincell{c}{Build the matrix of $\widehat H$ in this basis\\$\mathfrak H^{\rm CISD}_0= \left(\begin{array}{ccc} E_0^{\rm HF} & \langle \Phi_0 | \widehat H|\Phi_i^a\rangle & \langle \Phi_0| \widehat H | \Phi_{ij}^{ab}\rangle \\ * & \langle \Phi_i^a| \widehat H | \Phi_{j}^{b}\rangle & \langle \Phi_i^a| \widehat H | \Phi_{jk}^{bc}\rangle \\ * & * & \langle \Phi_{ij}^{ab}| \widehat H | \Phi_{kl}^{cd}\rangle \end{array}\right)$} & \tabincell{c}{Build the matrix of ${\rm Hess}_{\mathcal M}\mathcal{E}(x_0)$ in this basis\\$\mathfrak H_0^{\rm LR}=\left( \begin{array}{cc} \mathfrak h_{qq} & \mathfrak h_{qp} \\ \mathfrak h_{qp}^T & \mathfrak h_{pp} \end{array} \right)$}\\[35pt]\hline\hline
        Compute the eigenvalues of $\mathfrak H^{\rm CISD}_0$ & Compute the symplectic eigenvalues of $\mathfrak H^{\rm LR}_0$
    \end{tabular}}
\end{table}

\section{Dynamics on Grassmann manifolds (FCI, TDHF, TDDFT)}
\label{sec:Grassmann}

In this section, we restrict ourselves to the case when the K\"ahler manifold is the complex Grassmann manifold 
\begin{equation}\label{eq:Grassmann_manifold}
\mathcal M= {\rm Gr}_\C(r,n) = \left\{ \gamma \in  \C^{n \times n}_{\rm herm} \; | \; \gamma^2=\gamma, \; \tr(\gamma)=r \right\},
\end{equation}
where $r,n$ are positive integers such that $1 \le   r \le n-1$ (the trivial cases $\mathcal M= {\rm Gr}_\C(0,n)=\{0\}$ and  $\mathcal M= {\rm Gr}_\C(n,n)=\{I_n\}$ are uninteresting). Such manifolds appear in quantum dynamics in at least two settings:
\begin{enumerate}
\item pure-state dynamics of generic isolated finite-dimensional quantum systems described by the linear Schr\"odinger equation, or equivalently the linear quantum Liouville equation (see Section~\ref{subsec:linear_functionals}). In computational quantum chemistry, this corresponds to the FCI level of theory;\supercite{Knowles1984,Olsen1990} 
\item mean-field models for many-body fermionic systems. This framework encompasses the cases of both TDHF \supercite{Jrgensen1970, Olsen1988,Dalgaard1982,Matsuo2001,Ebata2010,Miura2010,vanMeer2017} and adiabatic TDDFT,\supercite{Runge1984,Gross1985, Onida2002,CASIDA1995,Casida1998,Casida2005} for which the dynamics is governed by a system of coupled nonlinear Schr\"odinger equations, or equivalently a nonlinear quantum Liouville equation (see Section~\ref{subsec:mean-field_models}).
\end{enumerate}
Before elaborating on these two fundamental examples, the geometry and K\"ahler structure of the complex Grassmann manifold ${\rm Gr}_\C(r,n)$ are recalled in Sections \ref{subsec:geo_Grassmann} and \ref{subsec:dyn_grass}, and the abstract Eqs.~\eqref{eq:dynamics}-\eqref{eq:linear_response} are rewritten in a more concrete form in the special case when $\mathcal M={\rm Gr}_\C(r,n)$ in Sections \ref{subsec:linear_functionals} and \ref{subsec:mean-field_models}. More details about Grassmann manifold can be found in the review articles~\cite{bendokat2024grassmann,edelman1998geometry} and the monographs~\cite{absil2008optimization,boumal2023introduction}.

\subsection{Geometry of complex Grassmann manifolds}
\label{subsec:geo_Grassmann}

The tangent space to the complex Grassmann manifold $\mathcal M= {\rm Gr}_\C(r,n)$ at some $\gamma \in \mathcal M$ is given by
$$
T_\gamma\mathcal M= \{Q \in \C^{n \times n}_{\rm herm} \; | \; \gamma Q + Q \gamma = Q \}.
$$
Any $\gamma \in \mathcal M$ can be decomposed as
$$
\gamma = U \left( \begin{array}{cc} I_r & 0 \\ 0 & 0 \end{array} \right) U^*, \quad \mbox{with} \quad UU^*=U^*U=I_{n},
$$
and we then have
\begin{equation}\label{eq:TgammaM}
T_\gamma\mathcal M= \left\{\left.Q=U \left( \begin{array}{cc} 0 & X \\ X^* & 0 \end{array} \right) U^*\right|X \in \C^{r \times m} \right\} \quad \mbox{with} \quad m:=n-r.
\end{equation}
The Grassmann manifold $\mathcal M$ can be endowed with the following K\"ahler structure:
\begin{itemize}
\item in view of Eq. \eqref{eq:TgammaM}, the tangent space $T_\gamma\mathcal M$ can be identified with the complex vector space $\C^{r \times m}$, endowed with the Hermitian inner product
$$
\forall \gamma \in \mathcal M, \quad \forall Q_1,Q_2 \in T_\gamma\mathcal M, \quad 
\langle Q_1,Q_2 \rangle_\gamma:= 2 \tr(X_1^*X_2) = 2 \tr\left( Q_1 \gamma Q_2 (1-\gamma) \right),
$$
where $X_j$ is the matrix of $\C^{r \times m}$ such that
\begin{equation} \label{eq:Grassmann_diag}
Q_j =  U \left( \begin{array}{cc} 0 & X_j \\ X_j^* & 0 \end{array} \right) U^*,\quad j=1,2;
\end{equation}
\item Riemannian metric: using the properties 
\begin{equation} \label{eq:prop_Gr}
    \gamma^2=\gamma, \quad (1-\gamma)^2=1-\gamma, \quad \forall Q \in T_\gamma\mathcal M_\R, \; \gamma Q+Q\gamma=Q  \mbox{ i.e., } \gamma Q = Q (1-\gamma),
\end{equation}
we get, for any $\gamma\in\mathcal{M}_{\R}$ and $Q_1,Q_2\in T_\gamma\mathcal{M}_{\R}$,
\begin{align*}
g_\gamma(Q_1,Q_2):&= {\rm Re}(\langle Q_1,Q_2 \rangle_\gamma) \\
&= \tr(X_1^*X_2)+\tr(X_1X_2^*) \\
&=  \tr\left( Q_1 \gamma Q_2 (1-\gamma) \right) +  \tr\left(  (1-\gamma) Q_2 \gamma Q_ 1 \right) \\
&= \tr\left( Q_1 \gamma Q_2 \right)+  \tr\left(  Q_1 (1-\gamma) Q_2 \right) \\
&= \tr(Q_1Q_2);
\end{align*}
\item Symplectic form: we infer from Eq. \eqref{eq:prop_Gr} that, for any $\gamma\in\mathcal{M}_{\R}$ and $Q_1,Q_2\in T_\gamma\mathcal{M}_{\R}$,
\begin{align*}
\omega_\gamma(Q_1,Q_2):&= {\rm Im}(\langle Q_1,Q_2 \rangle_\gamma) \\
&=  - i  \tr\left( Q_1 (2\gamma-1) Q_2 \right) \\ & = - i  \tr\left( Q_2 (1-2\gamma) Q_1 \right) .
\end{align*}
Note that, still using Eq.~\eqref{eq:prop_Gr}, we have
\begin{align*}
\overline{\tr(Q_1(2\gamma-1)Q_2)}&= \tr(Q_2(2\gamma-1)Q_1)= 2\tr(Q_2\gamma Q_1)-\tr(Q_2Q_1)\\
&=2\tr((1-\gamma)Q_2Q_1(1-\gamma))-\tr(Q_2Q_1) \\ 
&= \tr(Q_1(1-2\gamma)Q_2)= - \tr(Q_1(2\gamma-1)Q_2),
\end{align*}
so that $\tr(Q_1(2\gamma-1)Q_2)$ is an imaginary number.
\end{itemize}
A simple calculation shows that the symplectic operator $J_\gamma : T_\gamma \mathcal M_\R \to T_\gamma \mathcal M_\R$ is given by
\begin{equation} \label{eq:J_Grassmann}
\forall Q \in T_\gamma \mathcal M_\R, \quad J_\gamma Q := -i [Q,\gamma].
\end{equation}
In the representation \eqref{eq:Grassmann_diag}, it reads
$$
\forall Q = U  \left( \begin{array}{cc} 0 & X \\ X^* & 0 \end{array} \right) U^* \in T_\gamma\mathcal M_\R, \qquad  J_\gamma Q = U  \left( \begin{array}{cc} 0 & i X  \\ -i X^* & 0 \end{array} \right) U^*.
$$
Recall that the orthogonal projector $P_\gamma : \C^{n \times n}_{\rm herm} \to T_\gamma \mathcal M_\R$ (for the Frobenius inner product) has a simple expression:
\begin{equation}\label{eq:projector_Grassmann}
\forall \gamma \in \mathcal M_\R, \quad \forall M \in \C^{n \times n}_{\rm herm}, \quad P_\gamma M := \gamma M (1-\gamma) + (1-\gamma) M \gamma = [\gamma,[\gamma,M]],
\end{equation}
and that in the representation \eqref{eq:Grassmann_diag}, 
$$
\forall  M = U \left( \begin{array}{cc} A & B \\ B^* & C\end{array} \right) U^* \in \C^{n \times n}_{\rm herm}, \quad P_\gamma M =U \left( \begin{array}{cc} 0 & B \\ B^* & 0 \end{array} \right) U^* .
$$
It follows by direct calculations that
\begin{align} \label{eq:gamma_commutators}
\forall \gamma \in \mathcal M_\R, \quad \forall M \in \C^{n \times n}_{\rm herm}, \quad & i [\gamma,M] \in T_\gamma \mathcal M_\R, \quad  [\gamma,[\gamma,[\gamma,M]]]=[\gamma,M].
\end{align}
Finally, the exponential map is given by 
$$
\forall \gamma \in \mathcal M_\R, \quad \forall Q \in T_\gamma\mathcal M_\R, \quad {\rm Exp}_\gamma(Q) = e^{[\gamma,Q]} \gamma e^{-[\gamma,Q]},
$$
which in essence coincides with the exponential parametrization of molecular coefficients widely used in quantum chemistry. \supercite{helgaker2000molecular}

\subsection{Hamiltonian dynamics on Grassmann manifolds}
\label{subsec:dyn_grass}

Let $\mathcal M={\rm Gr}_\C(r,n)$ be a complex Grassmann manifold, $E: \C^{n \times n}_{\rm herm} \to \R$ an energy functional, smooth in a neighborhood of $\mathcal M_\R$, and $\mathcal E : \mathcal M_\R \to \R$ the restriction of $E$ to $\mathcal M_\R$. Denoting by $H_\gamma:=\nabla E(\gamma) \in \C^{n \times n}_{\rm herm}$ the gradient of $E$ at $\gamma \in \mathcal M_\R$ (for the Frobenius inner product), we have 
\begin{equation} \label{eq:proj_grad_grassmann}
{\rm grad}_{\mathcal M} \mathcal E(\gamma) = P_\gamma(\nabla E(\gamma)) = P_\gamma (H_\gamma) = [\gamma,[\gamma,H_\gamma]].
\end{equation}
Using Eqs.~\eqref{eq:J_Grassmann} and \eqref{eq:gamma_commutators}, we get
$$
\forall \gamma \in \mathcal M_\R, \quad  J_\gamma {\rm grad}_{\mathcal M} \mathcal E(\gamma) = -i [{\rm grad}_{\mathcal M}\mathcal E(\gamma),\gamma] = 
- i [[\gamma,[\gamma,H_\gamma]],\gamma] = -i [H_\gamma,\gamma].
$$
The symplectic dynamics~\eqref{eq:dynamics} generated by the energy functional $\mathcal E$ is thus given by
\begin{equation}\label{eq:dynamics_on_Grassmann}
{\frac{d\gamma}{dt}(t) = -i [H_{\gamma(t)},\gamma(t)] \quad \mbox{(K\"ahler dynamics on Grassmann manifolds)}.}
\end{equation}
Stationary points $\gamma_\star$ are characterized by the equation
\begin{equation} \label{eq:critical_point_Grassmann}
{[H_{\gamma_\star},\gamma_\star]=0} \quad \mbox{(critical points on Grassmann manifolds)}.
\end{equation}
This equilibrium condition implies that there exists a unitary matrix $U_\star \in U(n)$ such that 
\begin{equation}\label{eq:diag_H*}
\gamma_\star = U_\star \left( \begin{array}{cc} I_r & 0 \\ 0 & 0 \end{array} \right) U_\star^* \quad \mbox{and} \quad H_{\gamma_\star} = U_\star \left( \begin{array}{cc} E^r & 0 \\ 0 & E^m \end{array} \right) U_\star^*,
\end{equation}
where $E^r \in \C^{r \times r}_{\rm herm}$ and $E^m \in \C^{m \times m}_{\rm herm}$ are real diagonal matrices. 

\medskip

As $\mathcal M$ is a Riemannian submanifold of $\C^{n \times n}_{\rm herm}$, the Riemannian Hessian on $\mathcal E$ is given by 
$$
\forall Q \in T_{\gamma_\star} \mathcal M_\R, \quad {\rm Hess}_{\mathcal M} \mathcal E(\gamma_\star) (Q) = P_{\gamma_\star}( [D\mathcal G(\gamma_\star)](Q) ),
$$
where $\mathcal G: \C^{n \times n}_{\rm herm} \to \C^{n \times n}_{\rm herm}$ is any smooth extension of ${\rm grad}_{\mathcal M} \mathcal E(\gamma)$ in a neighborhood of~$\mathcal M_\R$. In view of~\eqref{eq:proj_grad_grassmann}, an obvious extension is
$$
\mathcal G(M) := [M,[M,\nabla E(M)]],
$$
for which 
$$
[D\mathcal G(\gamma_\star)](Q)=  \underbrace{[Q,[\gamma_\star,H_{\gamma_\star}]]}_{=0~\text{by Eq. \eqref{eq:critical_point_Grassmann}}} + \underbrace{[\gamma_\star,[Q,H_{\gamma_\star}]]}_{\in T_{\gamma_\star}\mathcal M_\R \; \mbox{\rm if} \; Q \in T_{\gamma_\star}\mathcal M_\R} +\underbrace{[\gamma_\star,[\gamma_\star,D^2E(\gamma_\star)(Q)]]}_{=P_{\gamma_\star}(D^2E(\gamma_\star)(Q))},
$$
where $D^2E(\gamma_\star):  \C^{n \times n}_{\rm herm} \to \C^{n \times n}_{\rm herm}$ is the second derivative of $E$ at $\gamma_\star$. The Riemannian Hessian ${\rm Hess}_{\mathcal M}\mathcal E(\gamma_\star) : T_{\gamma_\star} \mathcal M _\R \to T_{\gamma_\star} \mathcal M_\R$ is thus given by
\begin{equation} \label{eq:Hessian_Grassmann}
{\rm Hess}_{\mathcal M}\mathcal E(\gamma_\star) (Q) = [\gamma_\star,[Q,H_{\gamma_\star}]] +  P_{\gamma_\star} \left( D^2E(\gamma_\star) (Q) \right).
\end{equation}
Using Eqs. \eqref{eq:gamma_commutators} and \eqref{eq:critical_point_Grassmann}, we finally obtain that the linearized dynamics is given by
\begin{equation}\label{eq:linearized_dynamics_GM}
{\frac{dQ}{dt}(t) = -i [H_{\gamma_\star},Q] - i [D^2E(\gamma_\star)Q,\gamma_\star] \quad \mbox{(linear response on Grassmann manifolds)}}
\end{equation}
where we used the fact that Eqs. \eqref{eq:prop_Gr} and \eqref{eq:J_Grassmann} imply
$$
\forall Q \in T_{\gamma_\star}\mathcal M_\R, \quad J_{\gamma_\star} \left([ \gamma_\star,[Q,H_{\gamma_\star}]] \right) = -i [[ \gamma_\star,[Q,H_{\gamma_\star}]] ,\gamma_\star] = -i [H_{\gamma_\star},Q].
$$

\subsection{Linear energy functionals on projective spaces (FCI)}
\label{subsec:linear_functionals}

Consider an isolated quantum system with finite-dimensional state space $\mathcal H$.  In the quantum chemistry applications we have in mind, ${\mathcal H}$ is the FCI $N$-electron state-space generated by the $N_{\rm b}$ atomic spin-orbitals of the discretization basis, and is therefore of dimension $n=\binom{N_{\rm b}}{N}$. 

\medskip

The space $\mathcal H$ can be identified with $\C^n$ by the choice of an orthonormal basis. The postulates of quantum mechanics imply that the set of FCI pure states of the system can be identified with the projective space $P(\mathcal H) \cong {\rm CP}^{n-1}\cong {\rm Gr}(1,n)$. Recall that an element of ${\rm CP}^{n-1}$ can be represented by
\begin{enumerate}
\item a normalized vector $\psi \in \C^n$, it being understood that $\psi$ and $e^{i\theta}\psi$, with any $\theta \in \R$, represent the same element of ${\rm CP}^{n-1}$. This representation is the usual one in quantum mechanics (wavefunction formalism);
\item a non-zero vector $\widetilde\psi \in \C^n$, it being understood that $\widetilde\psi$ and $\eta \widetilde\psi$, with any $\eta \in \C\setminus\{0\}$, represent the same element of ${\rm CP}^{n-1}$;
\item a unique element $\Gamma \in {\rm Gr}_\C(1,n)$. This representation is used in the density-matrix formulation of quantum mechanics.
\end{enumerate}
The connection between these three representations is
$$
\Gamma = \psi\psi^* = \frac{\widetilde \psi \widetilde \psi^*}{|\widetilde\psi|^2} \quad \mbox{or, using Dirac's bra-ket notation,} \quad
\Gamma = |\psi\rangle \langle\psi| = \frac{|\widetilde \psi \rangle\langle \widetilde \psi|}{\langle \widetilde\psi|\widetilde\psi\rangle}.
$$
Let $H \in \C^{n \times n}_{\rm herm}$ be the Hamiltonian matrix of the system in the chosen orthonormal basis of the FCI $N$-electron state-space. Let $E: \C^{n \times n}_{\rm herm} \to \R$ be the linear energy functional defined by $E(\Gamma) = \tr(H\Gamma)$. The FCI energy functional is the restriction $\mathcal E$ of $E$ to ${\rm Gr}_\C(1,n)$:
$$
\mathcal E(\Gamma) = \tr(H\Gamma) = \langle \psi|H|\psi\rangle = \frac{\langle \widetilde\psi|H|\widetilde\psi\rangle}{\langle \widetilde\psi|\widetilde\psi\rangle}.
$$
The FCI approximation is thus a special case of the general framework described in Section~\ref{subsec:dyn_grass}, corresponding to $r=1$, $n=\binom{N_{\rm b}}{N}$, and $E: \C^{n \times n}_{\rm herm} \to \R$ linear. The dynamics  \eqref{eq:dynamics_on_Grassmann} then reads
\begin{equation}\label{eq:dynamics_Gr_1_n}
\frac{d\Gamma}{dt}(t) = -i [H,\Gamma(t)], \quad i\frac{d\psi}{dt}(t) = H\psi(t) - \beta(t) \psi(t), \quad i\frac{d\widetilde\psi}{dt}(t) = H\widetilde\psi(t) - \mu(t) \widetilde\psi(t),
\end{equation}
where $\beta(t) \in \R$, $\mu(t) \in \C$ are gauge parameters.
Steady states (see Eq.~\eqref{eq:critical_point_Grassmann}) are characterized by
$$
[H,\Gamma_\star]=0, \qquad \left\{ \begin{array}{l} H\psi_\star = E \psi_\star, \\ \|\psi_\star\|=1, \\ \psi_\theta(t) = e^{i\theta(t)} \psi_\star, \end{array} \right., \qquad \left\{ \begin{array}{l} H\widetilde\psi_\star= E \widetilde\psi_\star, \\ \widetilde\psi_\eta(t) = e^{\eta(t)} \psi_\star. \end{array} \right. 
$$
In particular, the local minimizers of $\mathcal E$ are global and are the ground-states of $H$ and the assumption that the ground-state $\Gamma_\star$ is nondegenerate is equivalent to the assumption that the ground-state eigenvalue $E_0$ of $H$ is simple. Other critical points of $\mathcal{E}$ are then exactly the excited states of $H$, corresponding to the energies $E_1\le E_2\le\cdots\le E_{n-1}$, and the excitation energies determined by CP are $\omega^{\rm CP-FCI}_k=E_k-E_0$ ($k=1,\ldots,n-1$).

\medskip

Let us turn to LR. Due to the fact that the energy functional $E(\Gamma)=\tr(H\Gamma)$ is linear in $\Gamma$, we have $H_{\Gamma_\star}=\nabla E(\Gamma_\star)=H$ and $D^2E(\Gamma_\star)=0$. It follows that the linear response Eq.~\eqref{eq:linearized_dynamics_GM} around a steady state $\Gamma_\star=|\Psi_\star\rangle\langle\Psi_\star|$ on ${\rm Gr}_\C(1,n)$ (with $H\Psi_\star=E_0 \Psi_\star$) reads
\begin{equation} \label{eq:LR-linear}
\frac{dQ}{dt}(t)=-i[H,Q(t)] \quad \mbox{on} \quad T_{\Gamma_\star} {\rm Gr}_\C(1,n)_\R.
\end{equation}
Using an orthonormal basis of the FCI Hamiltonian, in which the first basis vector is $\Psi_\star$, to identify the discretized $N$-particle state space with $\C^n$, $n=\binom{N_{\rm b}}{N}$, the representation in Eq.~\eqref{eq:diag_H*} for $r=1$ can be rewritten as
\begin{align*}
    &\Gamma_\star = \left( \begin{array}{cc} 1 & 0 \\ 0 & 0 \end{array} \right), \quad Q =  \left( \begin{array}{cc} 0 & \Psi^* \\ \Psi & 0 \end{array} \right), \quad \Psi\in\C^{n-1},\\
    & -i[H,Q]=  \left( \begin{array}{cc} 0 & (-i(H|_{\Psi_\star^\perp}-E_0)\Psi)^* \\ -i(H|_{\Psi_\star^\perp}-E_0)\Psi & 0 \end{array} \right) ,
\end{align*}
where $H|_{\Psi_\star^\perp}\in\C_{\rm herm}^{(n-1)\times (n-1)}$ denotes the projection of $H$ onto $\Psi_\star^\perp$. Consequently, Eq. \eqref{eq:LR-linear} is equivalent to
$$
\frac{d\Psi}{dt}(t) = -i(H|_{\Psi_\star^\perp}-E_0)\Psi(t) \quad \mbox{on} \quad \Psi_\star^\perp,
$$
which shows that the LR-FCI excitation energies are given as expected by $\omega^{\rm LR-FCI}_k=E_k-E_0$ ($k=1,\ldots,n-1$). 

In summary, we recover the well-known fact that both CP and LR are exact for linear quantum Hamiltonians (FCI level of theory in quantum chemistry).

\subsection{Mean-field models (TDHF and adiabatic TDDFT)} 
\label{subsec:mean-field_models}

Let us now consider the case of mean-field models. We will  focus on general HF (GHF) models, in which non-collinear spin states are allowed. However, the formalism presented here can be easily adapted to restricted and unrestricted HF models, more widely used in applications, as well as restricted/unrestricted adiabatic TDDFT.

\medskip

We still denote by $N_{\rm b}$ the number of atomic spin-orbitals of the chosen discretization basis, and assume that these functions are real-valued, which is the case for standard atomic orbital basis sets. We also assume that the system is not subjected to an external magnetic field and that spin-orbit coupling terms are neglected. In this case, it makes sense to seek HF or Kohn-Sham ground states satisfying time-reversal symmetry, leading to real-valued HF or Kohn-Sham occupied orbitals, and real-symmetric Fock or Kohn-Sham matrices at the ground state. In the rest of the section, we work in an orthonormal basis of the one-particle state-space $\mathcal H$ consisting of $N_{\rm b}$ real occupied/virtual molecular spin-orbitals $(\varphi_1,\ldots,\varphi_{N_b})$ at the GHF ground state $\gamma_\star$ with energies ranked in nondecreasing order. We denote by $h \in \C^{N_{\rm b} \times N_{\rm b}}_{\rm herm}$ the one-electron Hamiltonian matrix with
$$h_{pq}:=\int_{\R^3 \times \{ \alpha,\beta\}} \overline{\varphi_p(\boldsymbol{x})} \, \widehat h\varphi_q(\boldsymbol{x})\,{\rm d}\boldsymbol{x},\quad\text{where}~\widehat h~\text{is the one-body Hamiltonian,}$$
and by 
\begin{align*}
(pq|rs):= \int_{\R^3 \times \{ \alpha,\beta\}} \int_{\R^3 \times \{ \alpha,\beta\}} \frac{ \overline{\varphi_p(\boldsymbol x)}\varphi_q(\boldsymbol x) \overline{\varphi_r(\boldsymbol x')}\varphi_s(\boldsymbol x')}{|\boldsymbol{r}-\boldsymbol{r}'|}\,{\rm d}\boldsymbol{x}\,{\rm d}\boldsymbol{x}'
\end{align*}
the two-electron integrals in this basis. Here, $\boldsymbol{x}=(\boldsymbol{r}, \sigma)$ denotes the spatial and spin coordinates. Note that for the chosen basis, the complex conjugation in the above formula can be omitted as the basis functions are real. We also use the standard notation
$$
(pq||rs):= (pq|rs) - (ps|rq).
$$
The GHF energy functional $E^{\rm GHF}$ is given by
$$
E^{\rm GHF}(\gamma)={\rm Tr}(h\gamma)+\dfrac12{\rm Tr}(G(\gamma)\gamma),
$$
where $G:\C_{\rm herm}^{N_{\rm b}\times N_{\rm b}}\to\C_{\rm herm}^{N_{\rm b}\times N_{\rm b}}$ is defined as $G(\gamma)=J(\gamma)-K(\gamma)$, with $J,K:\C_{\rm herm}^{N_{\rm b}\times N_{\rm b}}\to\C_{\rm herm}^{N_{\rm b}\times N_{\rm b}}$ the Coulomb and exchange functionals, respectively:
\begin{equation}
    \label{eqn:HF Coulomb exchange term}
    [J(\gamma)]_{pq}:=\sum_{r,s=1}^{N_{\rm b}}(pq|rs)\gamma_{sr},\quad [K(\gamma)]_{pq}:=\sum_{r,s=1}^{N_{\rm b}}(ps|rq)\gamma_{sr}.
\end{equation}
It follows that $H_\gamma:=\nabla E^{\rm GHF}(\gamma)=h+G(\gamma)$ is the Fock matrix, that \eqref{eq:dynamics_on_Grassmann} coincides with the usual TDHF equations, and \eqref{eq:critical_point_Grassmann} with the McWeeny formulation of the HF equations.
In GHF theory, the no-unfilled shell property proved in Ref.~\cite{BLLS94} in the continuous setting also holds in the discretized setting. We therefore have that in the chosen basis of occupied/virtual real molecular spin-orbitals
\begin{align}
&\gamma_\star = \left( \begin{array}{cc} I_N & 0 \\ 0 & 0 \end{array} \right), \quad H_{\gamma_\star} = \left( \begin{array}{cc} {\rm diag}(\varepsilon_1,\ldots,\varepsilon_N) & 0 \\ 0 &  {\rm diag}(\varepsilon_{N+1},\ldots,\varepsilon_{N_{\rm b}}) \end{array} \right),  \label{eq:Aufbau} \\
& \mbox{with} \quad \varepsilon_1 \le \cdots \le  \varepsilon_N < \varepsilon_{N+1} \le \cdots \le \varepsilon_{N_{\rm b}}. \label{eq:NUS}
\end{align}

Let us now show that the K\"ahler formalism provide a simple alternative to Casida's derivation of LR excitation energies for mean-field models. Following the systematic strategy described in Section~\ref{sec:Kahler}, it indeed suffices to select a basis 
$$\left(u_1,\ldots,u_{N(N_{\rm b}-N)},-J_{\gamma_\star} u_1,\ldots,-J_{\gamma_\star} u_{N(N_{\rm b}-N)}\right)$$ 
of $T_{\gamma_\star}\mathcal M_\R$, build the matrix defined in Eq.~\eqref{eq:matrix_H*} with entries given by Eqs.~\eqref{eq:Hrr}-\eqref{eq:Hii}, and compute its symplectic eigenvalues. Choosing the canonical basis of $T_{\gamma_\star}\mathcal M_\R$, that is
$$
u_{[ia]}:=\frac{1}{\sqrt{2}}\left( \begin{array}{cc} 0 & e_{[ia]} \\ e_{[ia]}^T & 0 \end{array} \right), ~~ J_{\gamma_\star} u_{[ia]}
=\frac{1}{\sqrt{2}}\left( \begin{array}{cc} 0 & i e_{[ia]} \\ -i e_{[ia]}^T & 0 \end{array} \right), ~~ \mbox{with} ~~ [e_{[ia]}]_{jb} = \delta_{(ia),(jb)},
$$
for $1\le i,j\le N$ and $N+1\le a,b\le N_{\rm b}$, we deduce from Eq.~\eqref{eq:Hessian_Grassmann}, the fact that $D^2E^{\rm GHF}(\gamma)(Q) = G(Q)$, and the real character of the spin-orbitals that 
\begin{equation}
\mathfrak H_{\star}^{\rm LR-GHF} =  \left( \begin{array}{cc} A+B & 0 \\ 0 & A-B \end{array} \right), \label{eqn:H_symplectic}
\end{equation}
where the entries of the matrices $A$ and $B$ are given respectively by
$$
A_{ia,jb}:=(\varepsilon_a-\varepsilon_i)\delta_{ij}\delta_{ab}+ (ai||jb),\quad B_{ia,jb}:=(ai||bj).
$$
These matrices are the GHF analogues of the $A$ and $B$ matrices in LR-TDDFT Casida's equations. The linearized dynamics Hamiltonian is therefore separable and the matrix $\widetilde \Omega$ in Eq.~\eqref{eq:angular_frequencies_separable_Hamiltonians} then reads in this specific case
$$
\widetilde\Omega= \left((A+B)^{1/2}(A-B)(A+B)^{1/2} \right)^{1/2},
$$
an expression in agreement with Ref.~\cite[Eqs. (II.78)-(II.79)]{CASIDA20093}.

\medskip

Note that a more intrinsic way to write down the canonical basis of $T_{\gamma_\star}\mathcal M_\R$ is 
$$
H_{\gamma_\star}=\sum_{n=1}^{N_{\rm b}} \varepsilon_n |\varphi_n\rangle\langle\varphi_n|, \quad \langle\varphi_m|\varphi_n\rangle=\delta_{mn}, \quad \varepsilon_1 \le \cdots \le  \varepsilon_N < \varepsilon_{N+1} \le \cdots \le \varepsilon_{N_{\rm b}},
$$
$$
\gamma_\star= \sum_{i=1}^N |\varphi_i\rangle\langle\varphi_i|, \qquad T_{\gamma_\star}\mathcal M_\R = {\rm Span}_\R\left( E_{ia}, J_{\gamma_\star} E_{ia} \right),
$$
with
$$
 E_{ia}:=\frac{|\varphi_i\rangle\langle\varphi_a| + |\varphi_a\rangle\langle\varphi_i| }{\sqrt 2} \qquad \mbox{and} \qquad J_{\gamma_\star}E_{ia} = \frac{i |\varphi_i\rangle\langle\varphi_a| -i  |\varphi_a\rangle\langle\varphi_i| }{\sqrt 2}. 
$$

\section{Theoretical analysis in the weakly-interacting regime}
\label{sec:weakly_interacting}

\par In this part, we aim to compare theoretically, in the weakly-interacting regime, the CP and LR formalisms for computing the electronic excitation energies at the FCI and HF levels of theory. We consider a family of Hamiltonians parametrized by a coupling parameter $\eta$ that modulates the degree of interaction within the energy functional. In the non-interacting limit ($\eta=0$), HF and FCI lead to the same ground and excited state energies, so that CP-HF provides the same sequence of excitation energies as FCI, while LR-HF returns the subsequence of excitation energies corresponding to single excitations. Under a nondegeneracy assumption, we are able to compute the FCI (Section~\ref{subsec:theoretical comparison general framework}), as well as CP-HF and LR-HF (Section~\ref{subsec:CP-LR-HF}) excitation energies up to second-order errors in $\eta$ in the weakly-interacting regime ($|\eta| \ll 1$). To our knowledge, this is the first study aiming at comparing the CP and LR methodologies from a theoretical point of view. These theoretical findings are substantiated by numerical results in Section \ref{sec:numerics}.  We shall remark that although the cases of $\eta\ne1$ are non-physical, our analysis could offer potential insights into the study of more complicated real molecules.

In this section, we consider a system with $N=2 N_{\rm p}$ electrons ($N_{\rm p}$ is the number of electron pairs) in $N_{\rm b}$ atomic orbitals (hence $2N_{\rm b}$ atomic spin-orbitals) and the family of second-quantized Hamiltonians on the Fock space ${\rm Fock}(\C^{2N_{\rm b}})$ defined by
$$
\widehat H_\eta := \widehat h + \eta \widehat V = \sum_{\kappa,\lambda=1}^{N_{\rm b}} h_{\kappa\lambda} \sum_{\sigma \in \{\alpha,\beta\}} \widehat a_{\kappa\sigma}^\dagger \widehat a_{\lambda\sigma} + \frac{\eta}2 \sum_{\kappa,\lambda,\mu,\nu=1}^{N_{\rm b}} \sum_{\sigma,\sigma' \in \{\alpha,\beta\}} (\kappa\mu|\lambda\nu)_{\sigma\sigma'}  \widehat a_{\kappa\sigma}^\dagger \widehat a_{\lambda\sigma'}^\dagger  \widehat a_{\nu\sigma'} \widehat a_{\mu\sigma},
$$
where $\eta\ge0$ is a coupling parameter and the spin dependence of the integrals is indicated with the notation
\begin{align*}
(\kappa\mu||\lambda\nu)_{\sigma\sigma}:&=(\kappa_\sigma \mu_\sigma|\lambda_\sigma \nu_\sigma)-(\kappa_\sigma \nu_\sigma|\lambda_\sigma \mu_\sigma),~\sigma\in\{\alpha,\beta\},\\
(\kappa\mu|\lambda\nu)_{\sigma\sigma'}:&=(\kappa_\sigma \lambda_\sigma|\mu_{\sigma'} \nu_{\sigma'}),~\sigma \neq \sigma'\in\{\alpha,\beta\}.
\end{align*}
Let $(\phi_n^0)_{1\le n\le N_{\rm b}}$ be an orthonormal basis of $\C^{N_{\rm b}}$ diagonalizing the one-electron Hermitian matrix $h$, such that
$$
h\phi_n^0 =\varepsilon_n^0 \phi_n^0\quad\text{with}\quad\varepsilon_1^0\le\cdots\le\varepsilon_{N_{\rm b}}^0.
$$
To avoid technicalities, we assume that 
\begin{description}
\item[(H)] the eigenvalues of $h$ are nondegenerate and the $\binom{N_{\rm b}}{N_{\rm p}}$ real numbers obtained by summing up $N_{\rm p}$ of the $N_{\rm b}$ eigenvalues of $h$ are all different. 
\end{description}

\subsection{Analysis at the FCI level of theory}\label{subsec:theoretical comparison general framework}

It follows from Assumption (H) that the only degeneracies in the eigenstates of $\widehat H_0$ in the $N$-particle sector are due to degenerate spin states. Restricting our analysis to the $S_z=0$ spin sector, each state has $N_{\rm p}$ spin-up electrons and $N_{\rm p}$ spin-down electrons, the $N$-electron ground state of $\widehat H_0$ is a nondegenerate singlet state given by 
\begin{equation}
\Phi_0^{(0)}:= \frac{1}{\sqrt{N!}} \;  \phi_1^{\alpha,0}\wedge \phi_1^{\beta,0} \wedge \cdots \wedge \phi_{N_{\rm p}}^{\alpha,0} \wedge \phi_{N_{\rm p}}^{\beta,0},
\label{eqn:FCI ground state}
\end{equation}
where we adopt the notations 
$$\phi_n^{\sigma,0}:=\phi_n^0\otimes\sigma,\quad n=1,\ldots,N_{\rm b},\quad\sigma\in\{\alpha,\beta\},$$
and the first excited state is two-fold degenerate, spanned by
\begin{align*}
\Phi^{(0)}_\alpha &:= \frac{1}{\sqrt{N!}} \;  \phi_1^{\alpha,0} \wedge \phi_1^{\beta,0} \wedge \cdots \wedge \phi_{N_{\rm p}-1}^{\alpha,0} \wedge \phi_{N_{\rm p}-1}^{\beta,0} \wedge \phi_{N_{\rm p}+1}^{\alpha,0} \wedge \phi_{N_{\rm p}}^{\beta,0}, \\
\Phi^{(0)}_\beta &:= \frac{1}{\sqrt{N!}} \;  \phi_1^{\alpha,0} \wedge \phi_1^{\beta,0} \wedge \cdots \wedge \phi_{N_{\rm p}-1}^{\alpha,0} \wedge \phi_{N_{\rm p}-1}^{\beta,0}  \wedge \phi_{N_{\rm p}}^{\alpha,0} \wedge \phi_{N_{\rm p}+1}^{\beta,0}.
\end{align*}
Note that $\Phi^{(0)}_\alpha$ and $\Phi^{(0)}_\beta$ are not eigenfunctions of $\hat{S}^2$, but that 
$\Psi_{\rm S}^{(0)}:=(\Phi^{(0)}_\alpha-\Phi^{(0)}_\beta)/\sqrt 2$ is a singlet state (the first-excited state in the $\hat{S}^2=0$ sector) and $\Psi_{\rm T}^{(0)}:=(\Phi^{(0)}_\alpha+\Phi^{(0)}_\beta)/\sqrt 2$ the $S_z=0$ component of a triplet state (the ground state in the $\hat{S}^2=3/4$ sector). The energies of the ground state $\Phi^{(0)}_0$ and the two-fold degenerate first excited state are respectively
$$
E^{(0)}_0 = 2\sum_{i=1}^{N_{\rm p}} \varepsilon_i^0 \quad \mbox{and} \quad 
E^{(0)}_\alpha = E^{(0)}_\beta = E^{(0)}_{\rm S} = E^{(0)}_{\rm T}  = E^{(0)}_0 + \varepsilon_{N_{\rm p}+1} - \varepsilon_{N_{\rm p}}.
$$

\medskip

It follows from standard nondegenerate perturbation theory that for small $\eta$, the ground state of $\widehat H_\eta$ is nondegenerate and its energy $E_0(\eta)$ is a real-analytic function of $\eta$. Likewise the  first excited state in the $\hat{S}^2=0$ sector as well as the ground state in the $\hat{S}^2=3/4$ sector are nondegenerate, and their energies $E_{\rm S}(\eta)$ and $E_{\rm T}(\eta)$ are real-analytic functions of $\eta$ for $|\eta| \ll 1$. Consequently, so are the FCI excitation energies $\omega_{\rm S}^{\rm FCI}(\eta) := E_{\rm S}(\eta)- E_0(\eta)$ and $\omega_{\rm T}^{\rm FCI}(\eta) := E_{\rm T}(\eta)- E_0(\eta)$:
$$
\omega_{\rm S}^{\rm FCI}(\eta) = \sum_{k=0}^{+\infty} \eta^k \omega^{{\rm FCI},(k)}_{\rm S}, \quad \omega_{\rm T}^{\rm FCI}(\eta) = \sum_{k=0}^{+\infty} \eta^k \omega^{{\rm FCI},(k)}_{\rm T},
$$
where
$$
\omega^{{\rm FCI},(0)}_{\rm S}=\omega^{{\rm FCI},(0)}_{\rm T}= \varepsilon_{N_{\rm p}+1} - \varepsilon_{N_{\rm p}}.
$$
The coefficients $\omega^{(1)}_{\rm S}$ and $\omega^{(1)}_{\rm T}$ can be easily obtained by first-order perturbation theory:
\begin{align}
\omega^{{\rm FCI},(1)}_{\rm S/T}=&\sum_{j=1}^{N_{\rm p}-1}(N_{\rm p}+1,N_{\rm p}+1||j,j)_{\beta\beta}-\sum_{j=1}^{N_{\rm p}-1}(N_{\rm p},N_{\rm p}||j,j)_{\beta\beta} \nonumber \\
& +\sum_{j=1}^{N_{\rm p}}(N_{\rm p}+1,N_{\rm p}+1|j,j)_{\beta\alpha}\nonumber
-\sum_{j=1}^{N_{\rm p}}(N_{\rm p},N_{\rm p}|j,j)_{\beta\alpha} \\
& \pm (N_{\rm p},N_{\rm p}+1|N_{\rm p}+1, N_{\rm p})_{\beta\alpha}.\label{eqn:FCI_excitation_energy}
\end{align}
The first-order variation of the lowest excitation energy in the weakly-interacting regime ($0<\eta\ll1$) is thus given by the triplet state. 

\subsection{Analysis at the UHF level of theory}
\label{subsec:CP-LR-HF}

Let us now consider the UHF approximation, for which the underlying manifold is $\mathcal M={\rm Gr}_\C(N_{\rm p},N_{\rm b}) \times {\rm Gr}_\C(N_{\rm p},N_{\rm b})$. The UHF approximation of $\widehat H_\eta$ gives rise to a parametrized family of energy functionals of the form
\begin{align}
\label{eqn:HF energy functional}
 E^{\rm UHF}_\eta(\gamma_\alpha,\gamma_\beta) =& \tr(h\gamma_\alpha) + \tr(h\gamma_\beta) + \frac{\eta}2 \tr(J(\gamma_\alpha+\gamma_\beta)(\gamma_\alpha+\gamma_\beta)) \nonumber \\ & \quad -  \frac{\eta}2 \tr(K(\gamma_\alpha)\gamma_\alpha) -  \frac{\eta}2 \tr(K(\gamma_\beta)\gamma_\beta),
\end{align}
where $J$ and $K$ are defined in Eq. \eqref{eqn:HF Coulomb exchange term}.

\medskip

\noindent\textbf{CP-UHF.} The critical points of $\mathcal M \ni (\gamma_\alpha,\gamma_\beta) \mapsto \mathcal E^{\rm UHF}_0(\gamma_\alpha,\gamma_\beta)=\tr(h\gamma_\alpha) +\tr(h\gamma_\beta)   \in \R$ are the $(\gamma_{\alpha\star},\gamma_{\beta\star}) \in \mathcal M$ satisfying $[h,\gamma_{\alpha\star}]=[h,\gamma_{\beta\star}]=0$. Under the nondegeneracy assumption~(H), these critical points are isolated and there are exactly $\binom{N_{\rm b}}{N_{\rm p}} \times \binom{N_{\rm b}}{N_{\rm p}}$ of them. The ground state is 
\begin{equation}
\label{eqn:mean field ground state}
\left(\gamma_{[\emptyset,\emptyset]}^{(0)},\gamma_{[\emptyset,\emptyset]}^{(0)}\right) \quad \mbox{with} \quad \gamma_{[\emptyset,\emptyset]}^{(0)} = \sum_{j=1}^{N_{\rm p}} \phi_j^0 {\phi_j^0}^T,
\end{equation}
and the excited states are
\begin{equation}
\label{eqn:mean field excited states}
\left(\gamma_{\alpha,[I_\alpha,A_\alpha]}^{(0)},\gamma_{\beta,[I_\beta,A_\beta]}^{(0)} \right) \quad \mbox{where} \quad \gamma_{\sigma,[I_\sigma,A_\sigma]}:= \gamma_{[\emptyset,\emptyset]}^0 - \sum_{i \in I_\sigma} \phi_i^0{\phi_i^0}^T + \sum_{a \in A_\sigma} \phi_a^0{\phi_a^0}^T, 
\end{equation}
for $I_\sigma \subset \{1,\ldots,N_{\rm p}\}$, $A_\sigma \subset \{N_{\rm p}+1,\ldots,N_{\rm b}\}$, $|I|=|A| \ge 1$.
It follows from a simple application of the analytic implicit function theorem that under the nondegeneracy assumption (H), there exist $\eta_{\rm c} > 0$ and $\binom{N_{\rm b}}{N_{\rm p}} \times \binom{N_{\rm b}}{N_{\rm p}}$ real-analytic functions 
$$
\left(\gamma_{\alpha,[I_\alpha,A_\alpha]}(\bullet),\gamma_{\beta,[I_\beta,A_\beta]}(\bullet) \right)
:(-\eta_{\rm c},\eta_{\rm c}) \to \mathcal M, 
$$
$I_\alpha,I_\beta \subset \{1,\ldots,N_{\rm p}\}$, $A_\alpha,A_\beta \subset \{N_{\rm p}+1,\ldots,N_{\rm b}\}$, $|I_\alpha|=|A_\alpha|$, $|I_\beta|=|A_\beta|$,
such that for any $\eta \in (-\eta_{\rm c},\eta_{\rm c})$, the function $\mathcal M \ni (\gamma_\alpha,\gamma_\beta) \mapsto \mathcal E_\eta(\gamma_\alpha,\gamma_\beta) \in \R$ has exactly $\binom{N_{\rm b}}{N_{\rm p}} \times \binom{N_{\rm b}}{N_{\rm p}}$ critical points, all isolated, nondegenerate, and given by $\left(\gamma_{\alpha,[I_\alpha,A_\alpha]}(\eta),\gamma_{\beta,[I_\beta,A_\beta]}(\eta) \right)$. Note that, by symmetry, $\gamma_{\alpha,[\emptyset,\emptyset]}(\eta)=\gamma_{\beta,[\emptyset,\emptyset]}(\eta)=:\gamma_{[\emptyset,\emptyset]}(\eta)$, i.e., the UHF ground state coincides with the RHF ground state.

\medskip

For $\eta \in (-\eta_{\rm c},\eta_{\rm c})$, the excitation energies obtained by CP-UHF are thus the $\binom{N_{\rm b}}{N_{\rm p}} \times \binom{N_{\rm b}}{N_{\rm p}}-1$ positive numbers
\begin{equation} \label{eq:CP_HF}
\omega_{[I_\alpha,A_\alpha],[I_\beta,A_\beta]}^{{\rm CP-UHF}}(\eta) :=   E^{\rm UHF}_\eta\left(\gamma_{\alpha,[I_\alpha,A_\alpha]}(\eta),\gamma_{\beta,[I_\beta,A_\beta]}(\eta)\right) -  E^{\rm UHF}_\eta\left(\gamma_{[\emptyset,\emptyset]}(\eta),\gamma_{[\emptyset,\emptyset]}(\eta)\right)
\end{equation}
with $I_\sigma \subset \{1,\ldots,N_{\rm p}\}$, $A_\sigma \subset \{N_{\rm p}+1,\ldots,N_{\rm b}\}$, $|I_\sigma|=|A_\sigma|$, $|I_\alpha|+|I_\beta|=|A_\alpha|+|A_\beta| \ge 1$.
In addition, the functions $\omega_{[I_\alpha,A_\alpha],[I_\beta,A_\beta]}^{{\rm CP-UHF}}:(-\eta_{\rm c},\eta_{\rm c})\to \R$ are real-analytic and it holds
$$
 \omega_{[I_\alpha,A_\alpha],[I_\beta,A_\beta]}^{{\rm CP-UHF}}(\eta)=\sum_{k=0}^{+\infty} \eta^k \omega_{[I_\alpha,A_\alpha],[I_\beta,A_\beta]}^{{\rm CP-UHF},(k)},
 $$
 with
\begin{align}
\omega_{[I_\alpha,A_\alpha],[I_\beta,A_\beta]}^{{\rm CP-UHF},(0)}&=\sum_{a \in A_\alpha \cup A_\beta} \varepsilon_a^0 - \sum_{i\in I_\alpha \cup I_\beta} \varepsilon_i^0, \nonumber\\
\omega_{[I_\alpha,A_\alpha],[I_\beta,A_\beta]}^{{\rm CP-UHF},(1)}&=\left. \frac{\partial E^{\rm UHF}_\eta}{\partial \eta}\right|_{\eta=0}\left( \gamma_{\alpha,[I_\alpha,A_\alpha]}^{(0)},\gamma_{\beta,[I_\beta,A_\beta]}^{(0)} \right) - \left. \frac{\partial E^{\rm UHF}_\eta}{\partial \eta}\right|_{\eta=0}\left(  \gamma^{(0)}_{[\emptyset,\emptyset]} ,\gamma^{(0)}_{[\emptyset,\emptyset]} \right). \nonumber
\end{align}
The first-order variation of the excitation energy for the first excited state, given by the HOMO-LUMO excitation is 
\begin{align}
 \omega^{\text{CP-UHF},(1)}_{\rm HOMO-LUMO} 
 :=&
    \omega^{\text{CP-UHF},(1)}_{[\emptyset,\emptyset],[N_{\rm p},N_{\rm p}+1]} = \omega^{\text{CP-UHF},(1)}_{[N_{\rm p},N_{\rm p}+1],[\emptyset,\emptyset]} \nonumber  \\ =& \sum_{i=1}^{N_{\rm p}-1}(N_{\rm p}+1,N_{\rm p}+1||i,i)_{\sigma\sigma}+\sum_{i=1}^{N_{\rm p}} (N_{\rm p}+1,N_{\rm p}+1|i,i)_{\beta\alpha} \nonumber \\
    &-\sum_{i=1}^{N_{\rm p}-1}(N_{\rm p},N_{\rm p}||i,i)_{\sigma\sigma} -\sum_{i=1}^{N_{\rm p}} (N_{\rm p},N_{\rm p}|i,i)_{\beta\alpha}
    \label{eqn:UHF CP excitation energy first order}
\end{align}
for $\sigma=\alpha$ or $\beta$.

\medskip

\noindent\textbf{LR-UHF.} Let us now turn to the LR-UHF approximation of excitation energies. As the complex dimension of the manifold $\mathcal M$ is $2N_{\rm p}(N_{\rm b}-N_{\rm p})$, there are at most $2N_{\rm p}(N_{\rm b}-N_{\rm p})$ of them. For $\eta=0$, they can be interpreted as single excitations:
$$
\omega_{[i,a]}^{{\rm LR-UHF},(0)} :=\omega_{[i,a],[\emptyset,\emptyset]}^{{\rm LR-UHF},(0)} = \omega_{[\emptyset,\emptyset],[i,a]}^{{\rm LR-UHF},(0)} = \varepsilon_a^0-\varepsilon_i^0, \quad 1 \le i \le N_{\rm p}, \quad N_{\rm p}+1 \le a \le N_{\rm b}.
$$
In analogy with the FCI case, there are $N_{\rm p}(N_{\rm b}-N_{\rm p})$ two-fold degenerate excitation energies at $\eta=0$ due to the symmetry resulting from swapping spin-up and spin-down electrons. Likewise, the first-order variation of the lowest excitation energy is given by the triplet excited state as
\begin{align}
\omega_{\rm HOMO-LUMO}^{\rm LR-UHF,(1)}=&\sum_{j=1}^{N_{\rm p}-1}(N_{\rm p}+1,N_{\rm p}+1||j,j)_{\beta\beta}-\sum_{j=1}^{N_{\rm p}-1}(N_{\rm p},N_{\rm p}||j, j)_{\beta\beta}\nonumber\\
&+\sum_{j=1}^{N_{\rm p}}(N_{\rm p}+1,N_{\rm p}+1|j, j)_{\beta\alpha}-\sum_{j=1}^{N_{\rm p}}(N_{\rm p},N_{\rm p}|j, j)_{\beta\alpha}\nonumber\\
&-(N_{\rm p},N_{\rm p}+1|N_{\rm p}+1, N_{\rm p})_{\beta\alpha}.
\label{eqn:UHF LR excitation energy first order}
\end{align}
Notice that this expression coincides with the one found in the FCI case, in Eq.~\eqref{eqn:FCI_excitation_energy}. At first order, FCI thus coincides with LR in the UHF setting, both of which allow symmetry-breaking. However, when considering the restricted HF (RHF) case, for closed-shell systems, the expressions change and spin degeneracy is no longer an issue. For reference, we can construct a FCI in theory built only from Slater determinants with doubly occupied orbitals and will temporarily refer to it as restricted FCI here. In this case, restricted FCI can be shown to coincide with CP-RHF in the weakly interacting regime, but different from LR-RHF. Lastly, if we consider the general HF (GHF) in the spinless setting, it can be shown that all the three methods, FCI, LR-GHF, and CP-GHF coincide at $\eta=0$ up to the first order.

\section{Numerical simulations}
\label{sec:numerics}
We perform proof-of-concept numerical simulations to support our theoretical findings. In Section \ref{subsec:CP LR compare}, we compare the CP and LR formalisms in UHF theory for the H$_2$, H$_4$, and H$_2$O molecules, with a varying coupling parameter $\eta$. The exact FCI theory is taken as reference. In particular, we check the analytical first-order derivatives of the excitation energies at $\eta=0$ given by Eqs. \eqref{eqn:UHF CP excitation energy first order} and \eqref{eqn:UHF LR excitation energy first order} vs the numerical ones obtained by finite differences. In Section \ref{subsec:CP multiple solutions}, we demonstrate that multiple critical points can arise even for the H$_4$ molecule (rectangular geometry) at the UHF/STO-3G level of theory. The energies of the critical points and their overlaps with the exact FCI excited states are shown as functions of $\eta$. We note that in this section, only the index-1 saddle points (namely, the critical points where the Riemannian Hessian has exactly one negative eigenvalue) are considered in the CP formalism and they are computed numerically by using a constrained gentlest ascent dynamics, which is described in detail in a paper in preparation. The approximate excitation energies in the LR formalism are found by solving Casida's equations \supercite{CASIDA1995,Casida1998,Casida2005} using the open-source package PySCF.\supercite{sun2020recent}



\subsection{Comparison of CP and LR formalisms within the UHF theory}\label{subsec:CP LR compare}
In this subsection, numerical experiments performed in the UHF framework are shown. Calculations have been performed with the minimal basis set STO-3G\supercite{hehre1969a} and with the 3-21G basis,\supercite{binkley1980a} for the systems H$_2$, H$_4$ in a linear and a rectangular geometry, and H$_2$O, with $\eta$ in the range $[0,1]$. For H$_2$, a bond length of 1 a.u. was considered; for the H$_4$ molecule, the linear geometry considers a bond length of 0.875 {\AA} (Ref.~\cite{burton2020energy}); for the rectangular geometry, sides of 1 {\AA} and 2 {\AA} are considered. For the water molecule, the coordinates [O (0.00000000     0.00000000     0.11993333) a.u.], [H (0.00000000    -1.43497461    -0.95171452) a.u.], [H (0.00000000     1.43497461    -0.95171452) a.u.] were taken. 
We compare the first-order variation of the lowest excitation energies, obtained with both the LR and CP approaches.
We note that for the CP results in this subsection, only the lowest excited states at $\eta=0$, which correspond to the HOMO-LUMO excitations, have been tracked across the range of $\eta$. When $\eta$ increases, multiple critical points (with the same index) may emerge; see the discussions in Section \ref{subsec:CP multiple solutions}.

\begin{table}[!t]
    \centering
    \caption{First-order variations for the lowest excitation energies of the molecules H$_2$, H$_4$ in linear and rectangular geometry, and H$_2$O, obtained in the non-interacting regime of $\eta\xrightarrow[]{}0$. For both CP and LR, the results obtained by the analytic expressions Eqs.~\eqref{eqn:UHF CP excitation energy first order} and \eqref{eqn:UHF LR excitation energy first order}, denoted here by $\omega_{\rm anal}^{{\rm CP},(1)}$ and $\omega_{\rm anal}^{{\rm LR},(1)}$, and numerically using a finite-difference scheme, denoted here by $\omega_{\rm num}^{{\rm CP},(1)}$ and $\omega_{\rm num}^{{\rm LR},(1)}$, are listed. The difference $\Delta\omega_{\rm  anal}^{(1)}=\omega_{\rm anal}^{{\rm CP},(1)}-\omega_{\rm anal}^{{\rm LR},(1)}$ between the two approaches is also shown.}
    \label{tab:numerics_uhf}
    \resizebox{\linewidth}{!}{\begin{tabular}{|l|r|r|r|r|r|r|r|}
    \hline
         Molecule (Basis)& $\omega_{\rm num}^{{\rm FCI},(1)}$ & $\omega_{\rm anal}^{{\rm FCI},(1)}$ & $\omega_{\rm num}^{{\rm CP},(1)}$ & $\omega_{\rm anal}^{{\rm CP},(1)}$ & $\omega_{\rm num}^{{\rm LR},(1)}$ & $\omega_{\rm anal}^{{\rm LR},(1)}$ & $\Delta\omega_{\rm  anal}^{(1)}$  \\
         \hline
         H$_2$ (STO-3G)& -0.18283& -0.18283&-0.01259& -0.01259& -0.18283& -0.18283&0.17024\\
         H$_2$ (3-21G)& -0.42288& -0.42289& -0.33093& -0.33093& -0.42288& -0.42289& 0.09196\\
         H$_4$ lin. (STO-3G)& -0.24580& -0.24581& -0.10733& -0.10735& -0.24580& -0.24581& 0.13847 \\
         H$_4$ lin. (3-21G)&  -0.33247& -0.33249& -0.20751& -0.20753& -0.33247& -0.33249& 0.12496\\
         H$_4$ rect. (STO-3G)& -0.04026& -0.04025& 0.05212& 0.05212& -0.04027& -0.04025& 0.09236\\
         H$_4$ rect.(3-21G)& -0.13321& -0.13320& -0.06178& -0.06178& -0.13322& -0.13320& 0.07143\\
         H$_2$O (STO-3G)& -3.49415& -3.49721& -3.49414& -3.49415& -3.49720& -3.49721& 0.00306\\
         \hline
    \end{tabular}}
\end{table}

The results displayed in Table~\ref{tab:numerics_uhf} show for all examples a good agreement between the analytical derivatives obtained in Section~\ref{sec:weakly_interacting} and the numerical results obtained through a finite-difference scheme. The latter are calculated by
$$\omega_{\rm num}^{(1)} = \frac{\omega(\delta\eta)-\omega(0)}{\delta\eta}. $$ In these results, $\delta\eta$ has been chosen to be $10^{-4}$.
 Furthermore, we can observe that, for the examples considered here, the difference calculated as $\omega_{\mathrm{anal}}^{{\rm CP},(1)}-\omega_{\mathrm{anal}}^{{\rm LR},(1)}$ is always positive, showing that the LR result is always smaller than the CP one for the explored systems.

\begin{figure}[!t]
    \centering
    \includegraphics[width=0.5\linewidth]{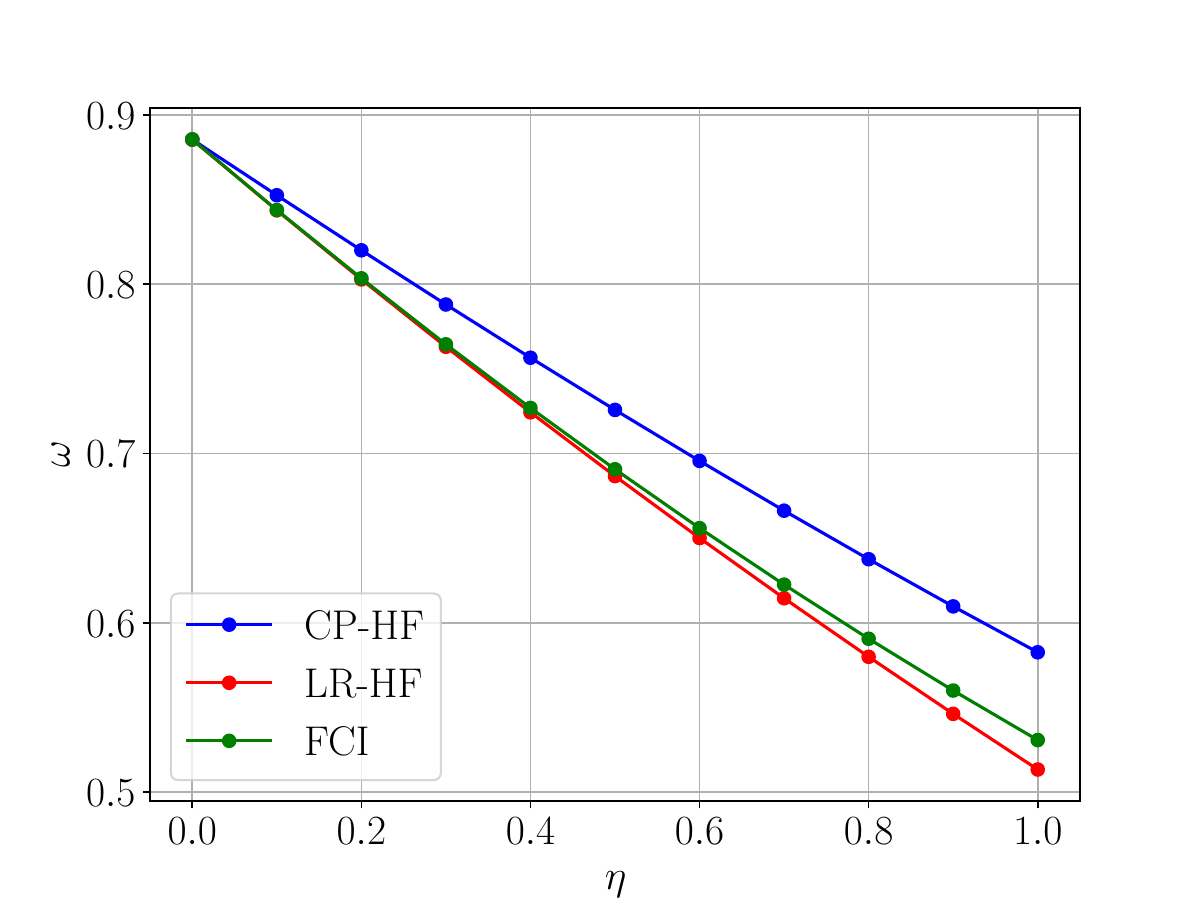}
    \caption{Dependence on the $\eta$-parameter of the excitation energy, calculated on the H$_2$ molecule, with the 3-21G basis set. }
    \label{fig:H2_0-1_uhf}
\end{figure}
\begin{figure}[!t]
    \centering
    \begin{subfigure}[t]{.49\linewidth}
        \centering
        \includegraphics[width=\linewidth]{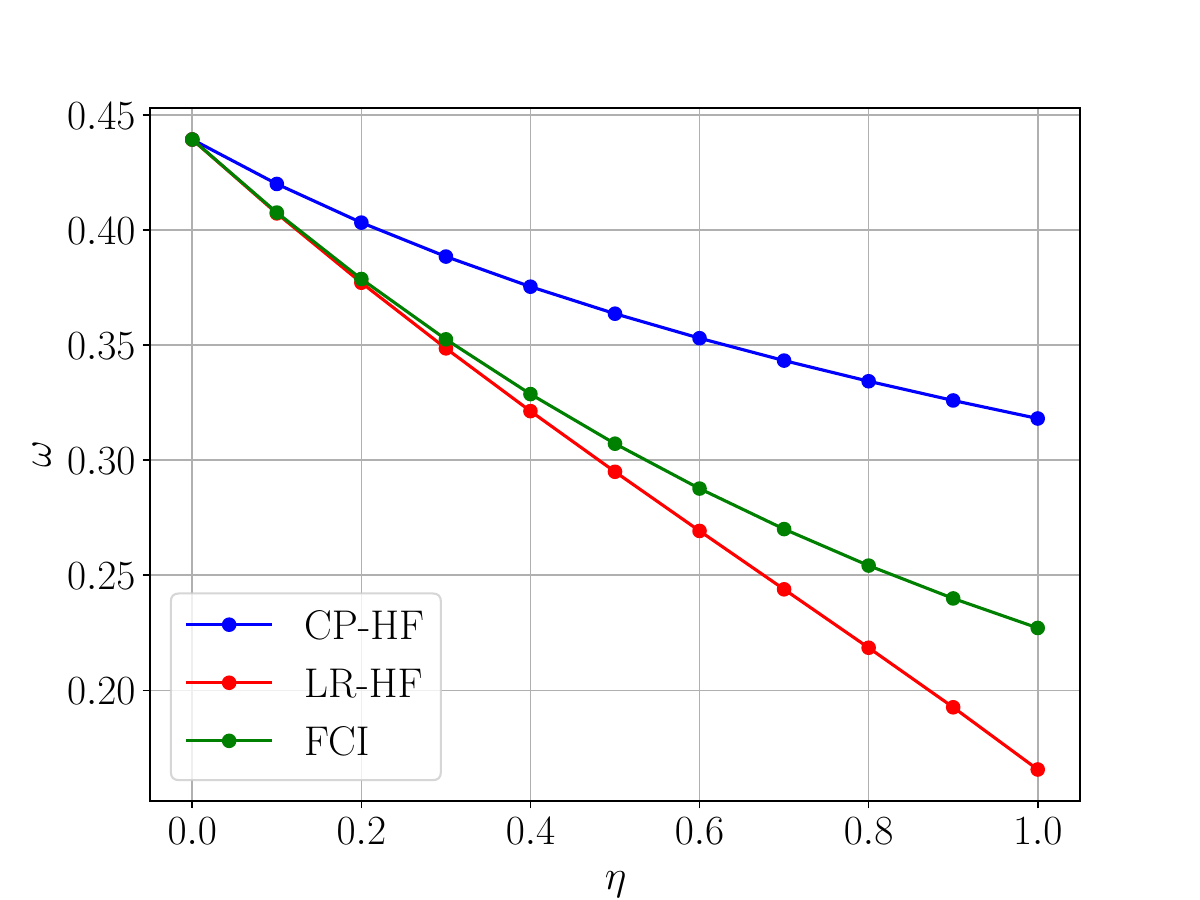}
        \caption{}
    \end{subfigure}
    \begin{subfigure}[t]{.49\linewidth}
        \centering
        \includegraphics[width=\linewidth]{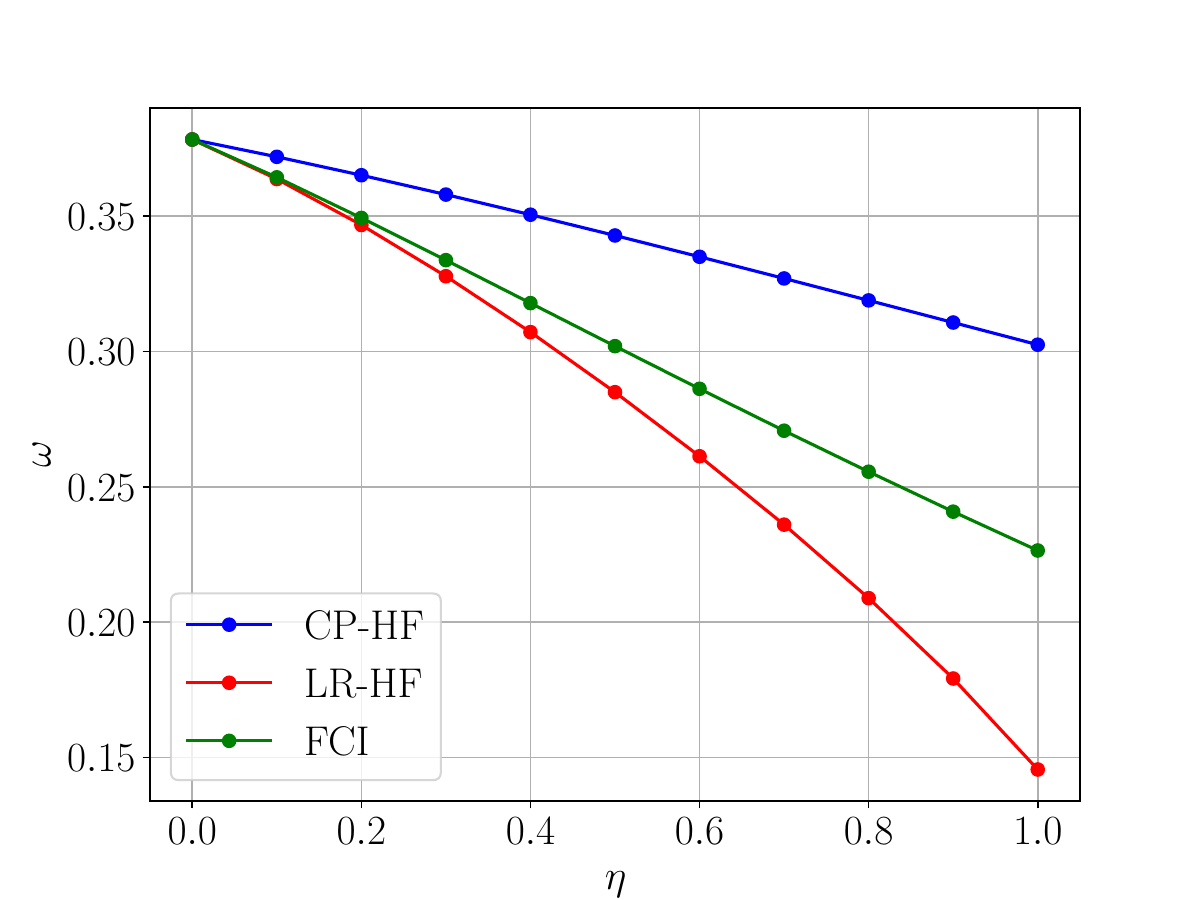}
        \caption{}
    \end{subfigure}
    
    \caption{Dependence on the $\eta$-parameter of the excitation energy, calculated on the H$_4$ molecule, in the linear (a) and rectangular (b) geometries, with the 3-21G basis set. }
    \label{fig:H4_lin_0-1_uhf}
\end{figure}

\begin{figure}[!t]
    \centering
    \begin{subfigure}[b]{0.49\textwidth}
        \centering
        \includegraphics[width=\textwidth]{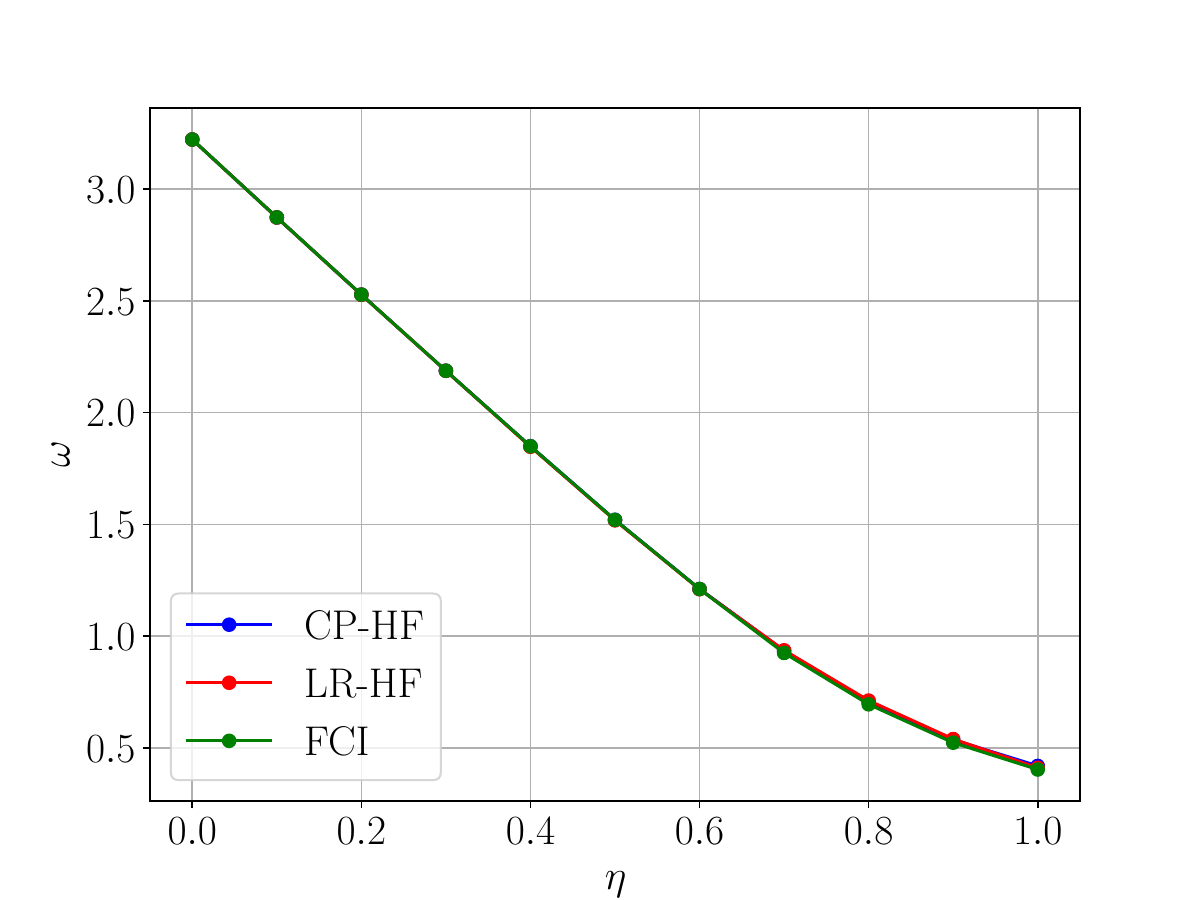}
        \caption{}
        \label{fig:H2O_0-1_tot}
    \end{subfigure}
    \begin{subfigure}[b]{0.49\textwidth}
        \centering
        \includegraphics[width=\textwidth]{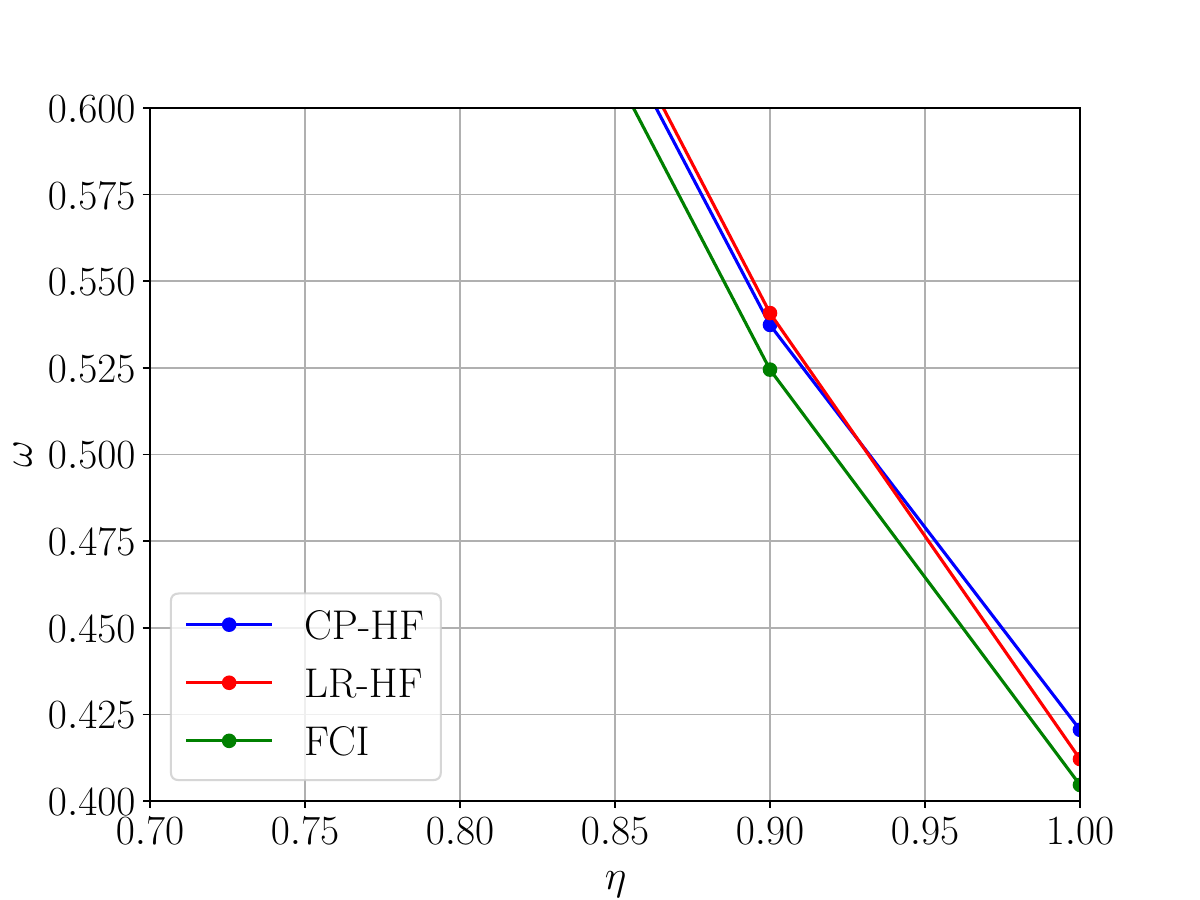}
        \caption{}
        \label{fig:H2O_0-1_ingrand}
    \end{subfigure}
    \caption{Dependence on the $\eta$-parameter of the excitation energy, calculated on the H$_2$O,  with the STO-3G basis set. In Fig.~\ref{fig:H2O_0-1_ingrand}, a zoom on the differences between the different approaches in the region around $\eta=1$ is plotted.}
    \label{fig:H2O_0-1_uhf}
\end{figure}

The two approaches CP and LR, at the UHF level of theory can be compared with the exact FCI results. The results are shown in Figs.~\ref{fig:H2_0-1_uhf}-\ref{fig:H2O_0-1_uhf}, for the molecules H$_2$ and H$_4$ in the linear and rectangular geometry, with the 3-21G basis set.\supercite{binkley1980a} 
\begin{figure}[!t]
    \centering
    \begin{subfigure}[b]{0.49\textwidth}
        \centering
        \includegraphics[width=\textwidth]{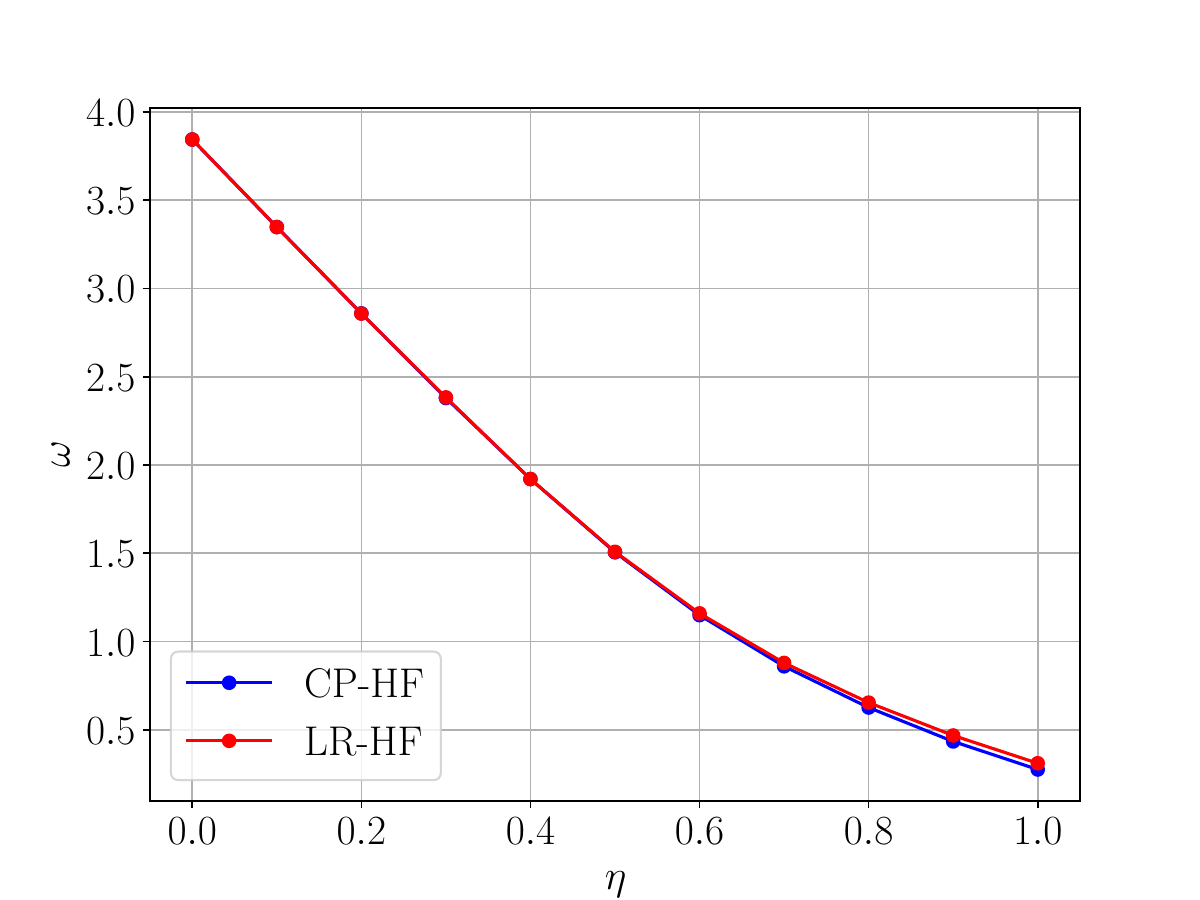}
        \caption{}
        \label{fig:H2O_0-1_6-31g}
    \end{subfigure}
    \begin{subfigure}[b]{0.49\textwidth}
        \centering
        \includegraphics[width=\textwidth]{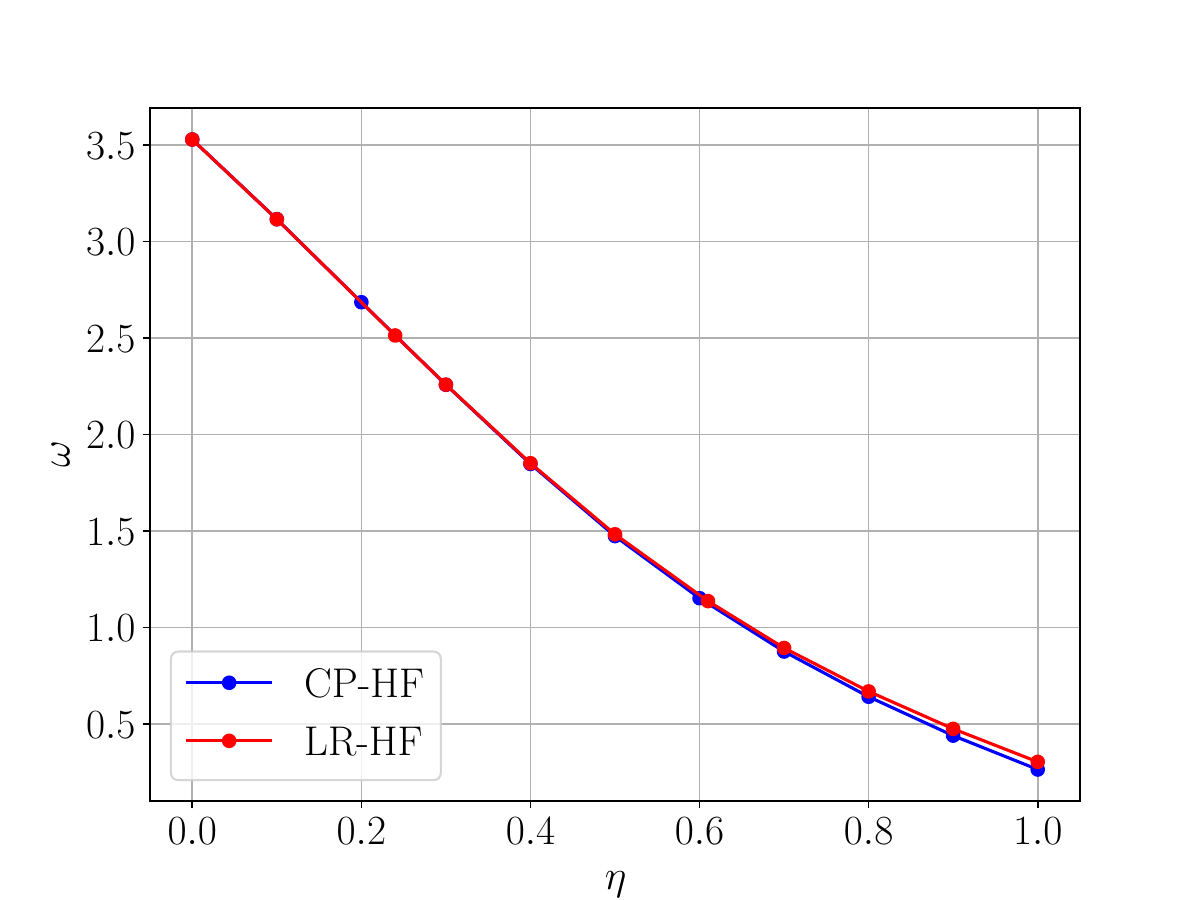}
        \caption{}
        \label{fig:H2O_0-1_ccpvdz}
    \end{subfigure}
    \caption{Dependence on the $\eta$-parameter of the excitation energy, calculated on H$_2$O. In Fig.~\ref{fig:H2O_0-1_6-31g} the 6-31G basis set is used, while in Fig.~\ref{fig:H2O_0-1_ccpvdz}, the cc-pVDZ basis set is used. At $\eta=0.2$ and $\eta=0.6$, the TD-HF algorithm in PySCF was not able to converge due to high linear dependencies.}
    \label{fig:H2O_0-1_uhf_larger_basis}
\end{figure}
For all cases, the LR results are closer to the FCI ones. The plots also show that, in the weakly-interacting limit ($\eta \to 0$), the LR formalism coincides with FCI at first order in $\eta$ as expected from Eqs. \eqref{eqn:FCI_excitation_energy} and \eqref{eqn:UHF LR excitation energy first order}, while the two approaches differ in the interacting case ($\eta=1$). In the water example with the STO-3G (Fig.~\ref{fig:H2O_0-1_uhf}),\supercite{hehre1969a} 6-31G,\supercite{ditchfield1971a,hehre1972a} and cc-pVDZ\supercite{dunning1989a} basis sets (Fig.~\ref{fig:H2O_0-1_uhf_larger_basis}), the three methods show small discrepancies across the range $\eta \in [0,1]$.

\subsection{Multiple critical points within the UHF theory for H$_4$}\label{subsec:CP multiple solutions}
Due to the nonlinearities induced by the UHF approximation, there can be multiple critical points (in particular, index-1 saddle points) over the UHF electronic landscape. In this subsection, we consider the H$_4$ molecule in a rectangular configuration with side lengths of 1 {\AA} and 1.1 {\AA}, at the UHF/STO-3G\supercite{hehre1969a} level of theory. The Hamiltonian is parametrized by the coupling parameter $\eta$, as in Section~\ref{subsec:theoretical comparison general framework}. For each $\eta\in[0,2]$ with a spacing of $0.05$, we find the index-1 saddle points by running the saddle search algorithm from 50 different randomly generated initial points over the Grassmannian.

\begin{figure}[]
    \centering
    \includegraphics[width=.5\textwidth]{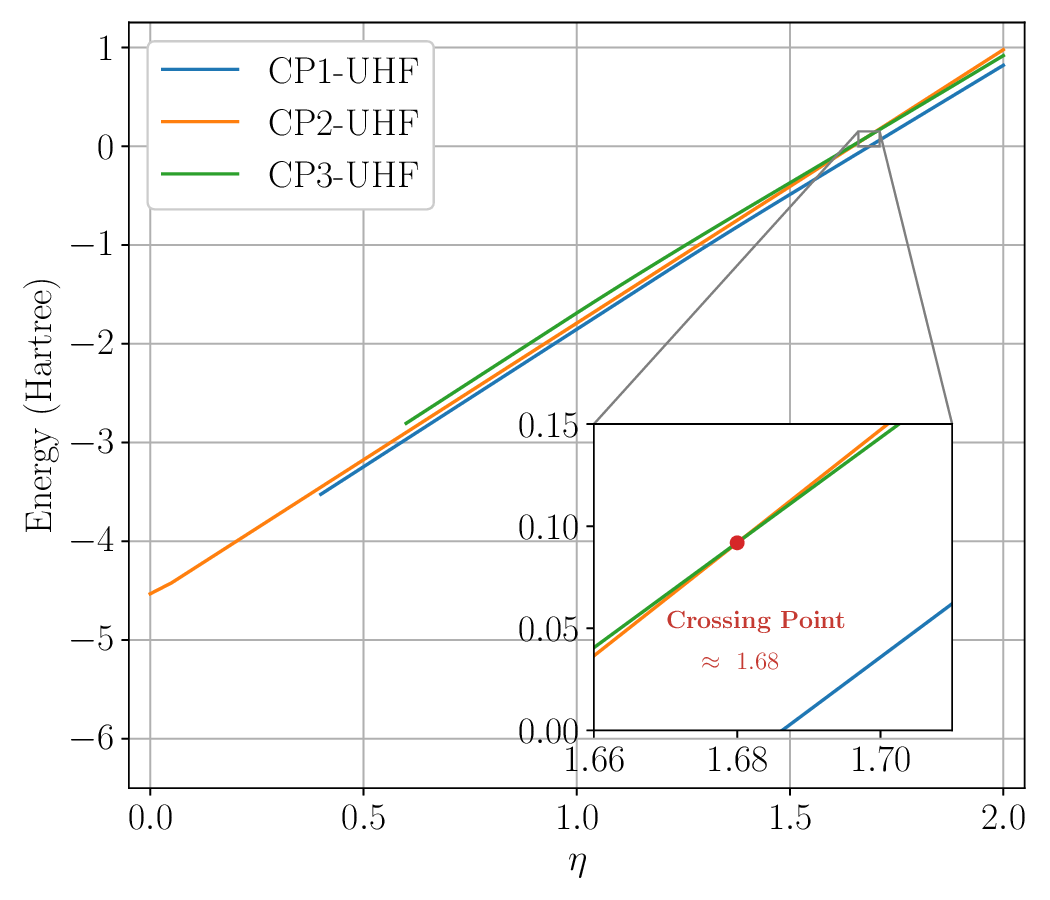}
    \caption{Dependence on the $\eta$-parameter of the energy of the index-1 saddle points, calculated on H$_4$ in a rectangular configuration with sides of 1 \AA \;and 1.1 \AA. The UHF/STO-3G level of theory has been used. Calculations have been made in steps of $\Delta \eta=0.2$.}
    \label{fig:H4_rect_crossing}
\end{figure}

Fig.~\ref{fig:H4_rect_crossing} shows the energies of the index-1 saddle points which appear in the range $\eta \in [0,2]$. 
We see that in this range, at most three critical points are found. In the range $\eta\in[0,0.35]$, only one index-1 saddle point is found, in $\eta\in[0.40,0.60]$ two index-1 saddle points are found, and from there with increasing $\eta$, three index-1 saddle points are observed. At $\eta\approx1.68$ the lines corresponding to the second and third saddle points are crossing, showing that the ordering in the energies of the states might not be preserved.

\begin{figure}[!t]
    \centering
    \begin{subfigure}[t]{0.49\textwidth}
        \centering
        \includegraphics[width=\linewidth]{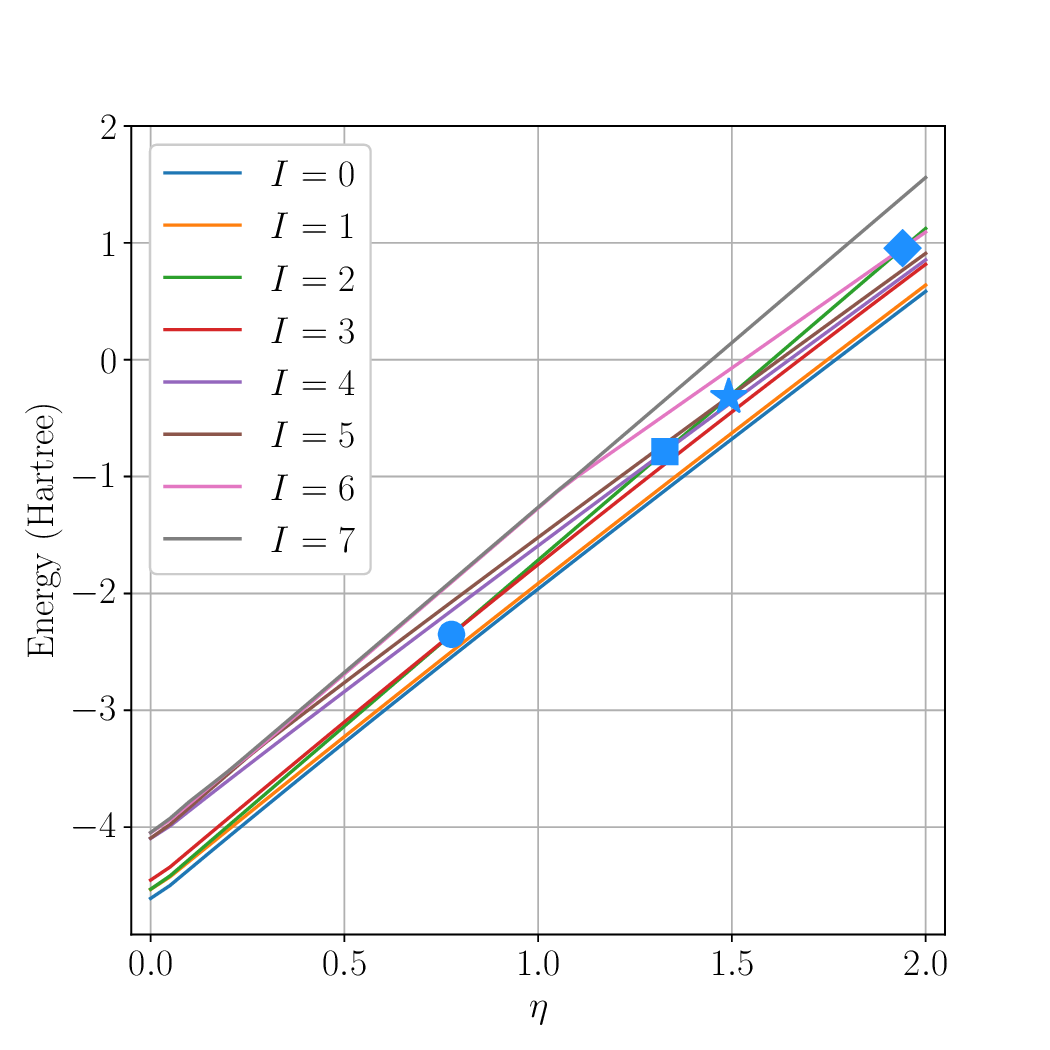}
        \caption{}
        \label{fig:fci_en}
    \end{subfigure}
    \begin{subfigure}[t]{0.49\textwidth}
        \centering
        \includegraphics[width=\linewidth]{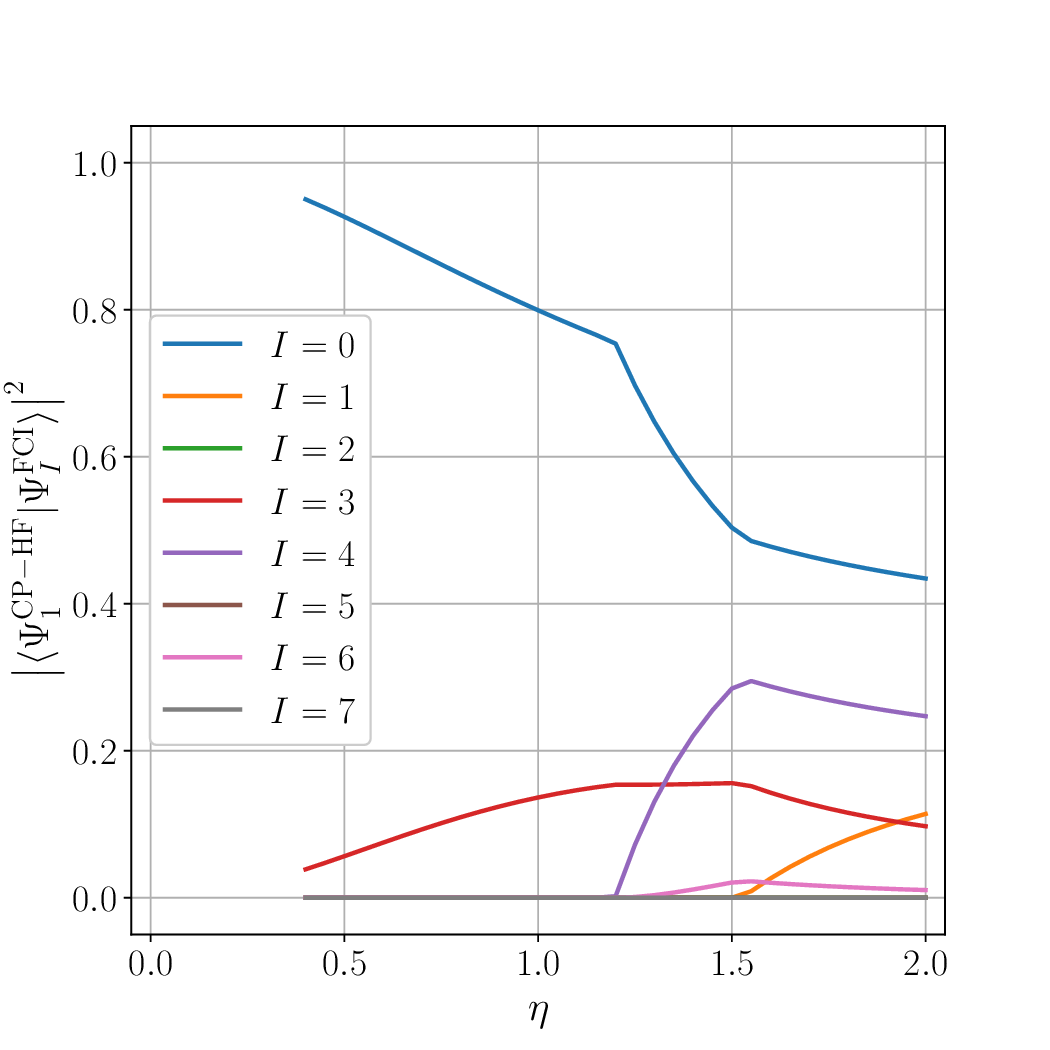}
        \caption{}
        \label{fig:H4_CP1}
    \end{subfigure}\\
    \begin{subfigure}[t]{0.49\textwidth}
        \centering
        \includegraphics[width=\linewidth]{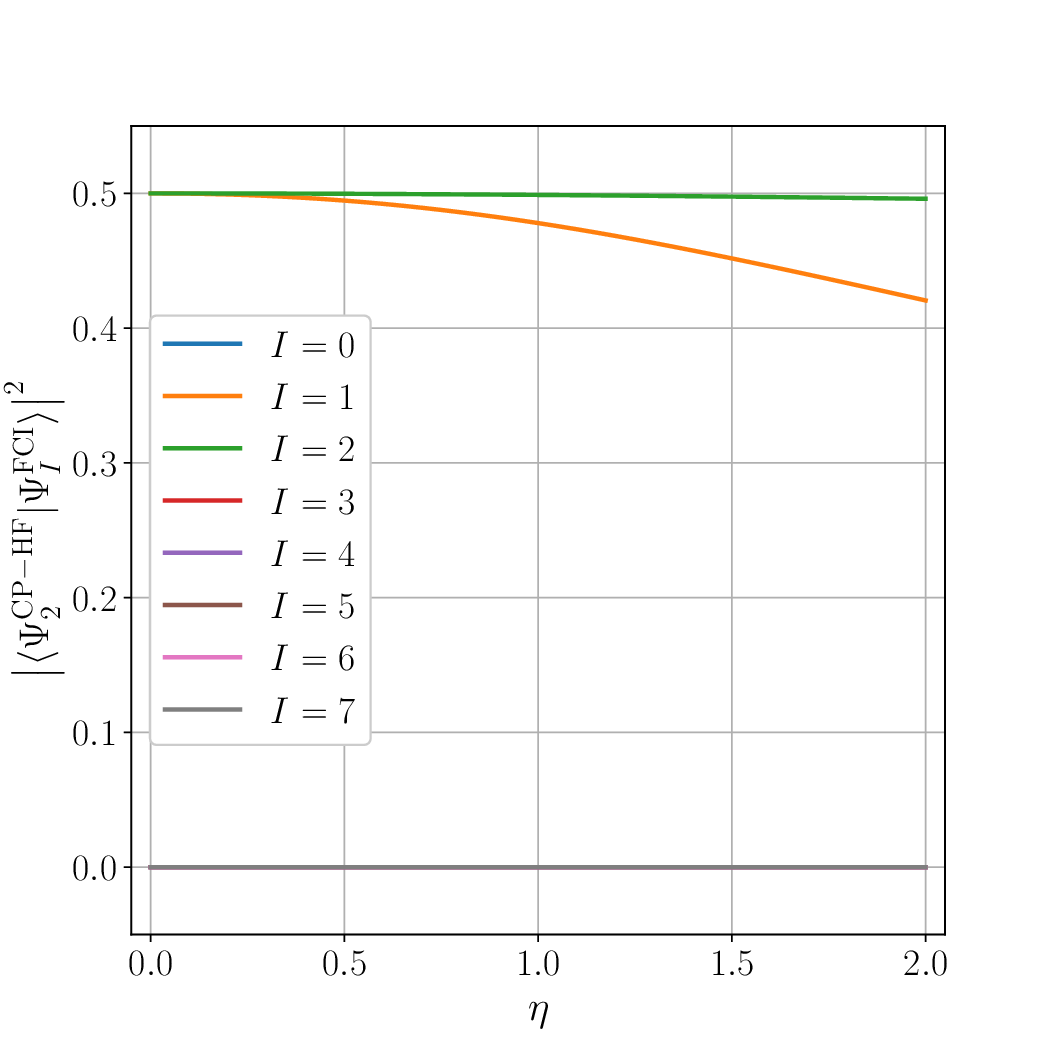}
        \caption{}
        \label{fig:H4_CP2}
    \end{subfigure}
    \begin{subfigure}[t]{0.49\textwidth}
        \centering
        \includegraphics[width=\linewidth]{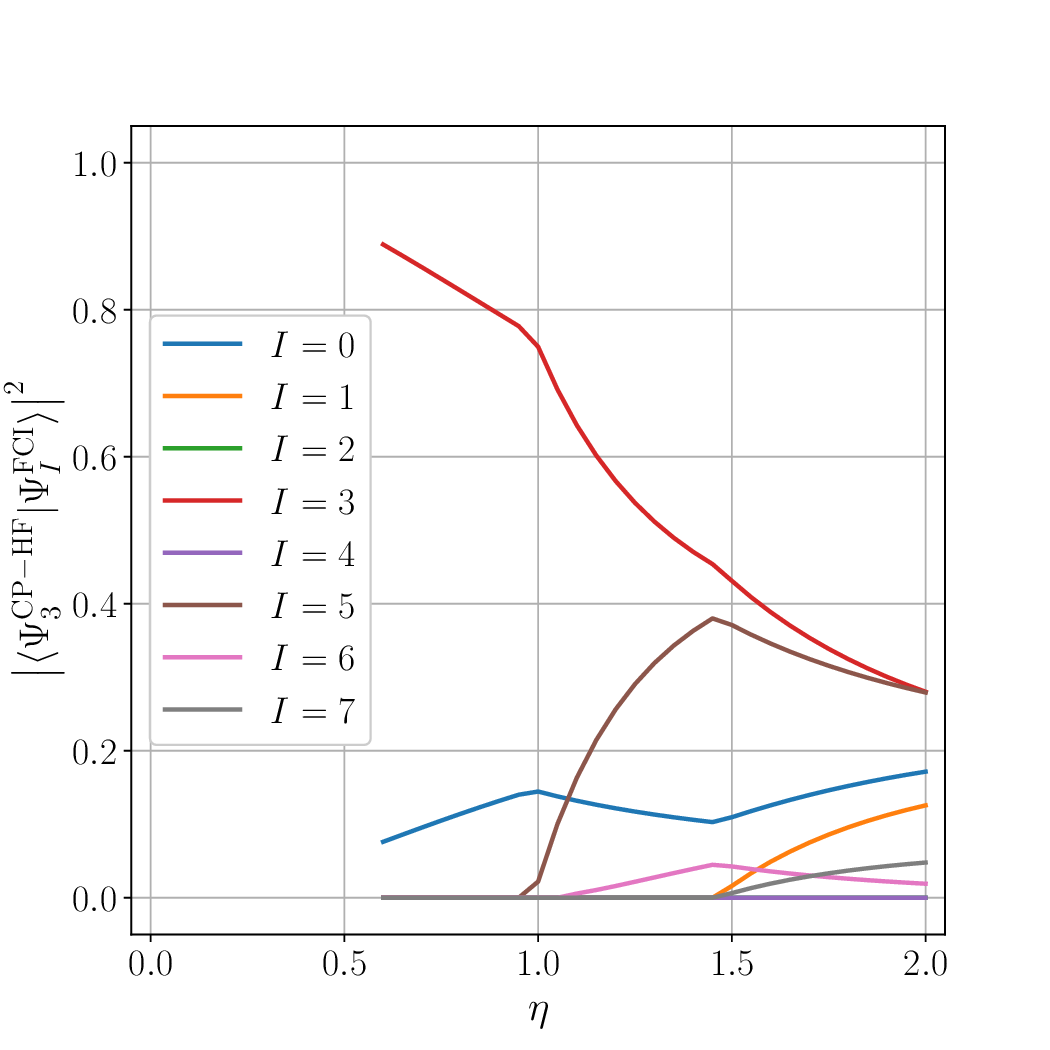}
        \caption{}
        \label{fig:H4_CP3}
    \end{subfigure}
    \caption{(a) Dependence on the $\eta$-parameter of the FCI energies, and (b-d) dependence on the $\eta$-parameter of the overlaps of the UHF critical points with the FCI states, calculated on H$_4$ in a rectangular configuration with sides of 1 \AA~and 1.1 \AA. Calculations have been made in steps of $\Delta\eta=0.05$.}
    \label{fig:overlaps_CP_fci}
\end{figure}

It is known that not all index-1 saddle points provide suitable approximations for excited states. An interpretation of the critical points can be gained through projection onto the FCI states, as these form a complete basis. Our calculations consider the case of $S_z=0$, therefore fixing $N_\alpha=N_\beta=2$. In the present case of a minimal basis set, a total number of 36 FCI states can be built in the $S_z=0$ sector of the FCI space. The projection of the wave functions corresponding to the critical points can be written as
\[\ket{\Psi^{\mathrm{CP-HF}}(\eta)}=\sum_{I}c_I(\eta)\ket{\Psi_I^{\mathrm{FCI}}(\eta)}.\]
The magnitudes of the $c_I(\eta)$ coefficients indicate the extent to which the critical points correspond to the $I$-th FCI state. For the sake of clarity, we will include only the eight lowest-lying FCI states, as the contributions of the higher-energy states are relatively small. We include the evolution of the considered FCI excited energies with the coupling parameter in Fig.~\ref{fig:fci_en}. 
We label these states as $\Psi_0^{\mathrm{FCI}}(\eta)$, $\Psi_1^{\mathrm{FCI}}(\eta)$, $\ldots$, $\Psi_7^{\mathrm{FCI}}(\eta)$
in accordance to their energy ordering in the non-interacting case. At $\eta=0$, we observe that the pairs ($\Psi_1^{\mathrm{FCI}}(0)$, $\Psi_2^{\mathrm{FCI}}(0)$), ($\Psi_4^{\mathrm{FCI}}(0)$, $\Psi_5^{\mathrm{FCI}}(0)$), and ($\Psi_6^{\mathrm{FCI}}(0)$, $\Psi_7^{\mathrm{FCI}}(0)$) are degenerate. These degeneracies can be understood by considering the occupation of the molecular orbitals. In the non-interacting case, the lowest eight FCI states are either singlets or triplets and are therefore either single Slater determinants or linear combinations of two Slater determinants:
\begin{align*}
    \ket{\Psi_0^{\rm FCI}(0)}&=\ket{\phi_1^{\alpha,0}\phi_1^{\beta,0}\phi_2^{\alpha,0}\phi_2^{\beta,0}}\\
    (\ket{\Psi_1^{\rm FCI}(0)},\ket{\Psi_2^{\rm FCI}(0)})&=\frac{1}{\sqrt{2}}(\ket{\phi_1^{\alpha,0}\phi_1^{\beta,0}\phi_2^{\alpha,0}\phi_3^{\beta,0}}\pm\ket{\phi_1^{\alpha,0}\phi_1^{\beta,0}\phi_2^{\beta,0}\phi_3^{\alpha,0}})\\
    \ket{\Psi_3^{\rm FCI}(0)}&=\ket{\phi_1^{\alpha,0}\phi_1^{\beta,0}\phi_3^{\alpha,0}\phi_3^{\beta,0}}\\
    (\ket{\Psi_4^{\rm FCI}(0)},\ket{\Psi_5^{\rm FCI}(0)})&=\frac{1}{\sqrt{2}}(\ket{\phi_1^{\alpha,0}\phi_2^{\beta,0}\phi_2^{\alpha,0}\phi_3^{\beta,0}}\pm\ket{\phi_1^{\beta,0}\phi_2^{\beta,0}\phi_2^{\alpha,0}\phi_3^{\alpha,0}})\\
    (\ket{\Psi_6^{\rm FCI}(0)},\ket{\Psi_7^{\rm FCI}(0)})&=\frac{1}{\sqrt{2}}(\ket{\phi_1^{\alpha,0}\phi_1^{\beta,0}\phi_2^{\alpha,0}\phi_4^{\beta,0}}\pm\ket{\phi_1^{\alpha,0}\phi_1^{\beta,0}\phi_2^{\beta,0}\phi_4^{\alpha,0}})\\
\end{align*}
The degeneracies arise from the fact that at $\eta=0$, the singlet state and the triplet state of $S_z=0$ have the same energy, as discussed in Section~\ref{subsec:theoretical comparison general framework}.

For $\eta>0$, it can be observed that the degeneracy is lifted, as the singlet and triplet states no longer give the same energy. With increasing interaction parameter $\eta$, the FCI states are observed to cross. In particular, $\Psi_2^{\mathrm{FCI}}$ (green line in Fig.~\ref{fig:fci_en}) crosses consecutively with the states $\Psi_3^{\mathrm{FCI}}$, $\Psi_4^{\mathrm{FCI}}$, $\Psi_5^{\mathrm{FCI}}$, $\Psi_6^{\mathrm{FCI}}$ (red, purple, brown, and pink lines in Fig.~\ref{fig:fci_en}) at $\eta\approx0.77$, 1.33, 1.49, and 1.94 respectively, as indicated by the dot, square, star, and diamond markers in the figure.

In Figs.~\ref{fig:H4_CP1}-\ref{fig:H4_CP3}, the $|c_I|^2$ weights corresponding to the three index-1 saddle points are plotted. In Fig.~\ref{fig:H4_CP1}, the analysis is performed for the lowest-lying critical point (CP1, blue line in Fig.~\ref{fig:H4_rect_crossing}). This saddle point appears only from $\eta\approx0.4$ and shows to be dominated by the FCI ground state. Therefore, this critical point can be interpreted as a spurious saddle point arising from the use of an approximate wave function method (UHF), which introduces nonlinearities. The existence of spurious critical points in nonlinear approximation models has been reported many times since the early days of quantum chemistry, and the study of this phenomenon is still an active field of research (cf. e.g., the recent work by Marie and Burton in Ref.~\cite{Marie2023} on CASSCF). 
Fig.~\ref{fig:H4_CP2} shows the same analysis for the second critical point (CP2, orange line in Fig.~\ref{fig:H4_rect_crossing}). This state remains a combination of $\Psi_1^{\rm FCI}$ and $\Psi_2^{\rm FCI}$. In particular, at $\eta=0$, we have $\ket{\Psi_{2}^{\mathrm{HF,CP}}(0)}=\frac{1}{\sqrt{2}}(\ket{\Psi_1^{\mathrm{FCI}}(0)}+\ket{\Psi_2^{\mathrm{FCI}}(0)})$ at $\eta=0$. As the two states $\Psi_1^{\rm FCI}$ and $\Psi_2^{\rm FCI}$ are degenerate at $\eta=0$ but belong to different spin-symmetry sectors, this critical point is a symmetry-broken solution, formed by a linear combination of the singlet first excited state $\Psi_1^{\rm FCI}$ and the  triplet ground state $\Psi_2^{\rm FCI}$. In Fig.~\ref{fig:H4_CP3}, the overlap of the third critical point (CP3, green line in Fig.~\ref{fig:H4_rect_crossing}) is shown. It is observed that this critical point, which arises only from $\eta=0.60$, has a large overlap with the state $\Psi_3^{\mathrm{FCI}}$. In particular, for $\eta\in[0.60,1]$ the overlap is very close to one, but then the saddle point becomes a linear combination of states $\Psi_3^{\mathrm{FCI}}$ and $\Psi_5^{\mathrm{FCI}}$. 

The analysis confirms once more that the interpretation of critical points as excited states should therefore be performed with care. On the reported example, only one critical point of a Morse index 1 (CP2 in Fig.~\ref{fig:H4_rect_crossing}) can be interpreted as a proper approximation of the first excited state in the UHF setting, while the other two are spurious states arising from the intrinsic nonlinearity of the UHF theory.

\section{Conclusions}
In this work, a unified and compact framework, based on the concept of Kähler manifolds, has been presented for the description of excited states in electronic structure theory, covering two different approaches, critical point search (CP) and linear response theory (LR). In particular, we describe the framework in a pedagogical way over the Grassmann manifolds, which amounts to be a more straightforward and systematic formulation of LR than Casida's derivations when dealing with the mean-field models. 

It was shown that CP and LR coincide for the FCI level of theory, but give different formulations for nonlinear approximations of the wave function. 
For the mean-field models such as HF and DFT, the CP approach amounts to finding different solutions of the HF or the KS equations, while the LR approach boils down to solving Casida's equations, commonly used in LR-TDHF and LR-TDDFT.
We have pointed out that TDHF and TDDFT can only target single excitations. On the contrary, the CP approach can target higher excitations and allows for orbital relaxation, at the cost of introducing additional or even non-physical candidates for excited states.

\par To analyze the differences between the CP and LR approaches, we have explored the weakly-interacting regime, based at the non-interacting limit, where the energy functional becomes linear and the LR and CP approaches are equivalent to compute the lowest excitation energies regardless of the level of theory in use. In particular, we exploit the perturbation theory to find the analytic first-order derivatives of excitation energies with respect to an additional coupling parameter, controlling the amount of interaction. For example in the UHF level of theory, we find that the CP approach yields a derivative that differs from the one in the FCI level, while the LR approach gives the correct first-order expression for the weakly-interacting case. 

Preliminary numerical simulations have been performed to show the comparison of the CP and LR approaches in the UHF level of theory as well as the FCI, on the chosen systems H$_2$, H$_4$, and H$_2$O, focusing on the first excited state. For the CP approach, the search is restricted to index-1 saddle points. The simulations have confirmed the analytic expressions for the first-order excitation energies at the non-interacting regime. Furthermore, we have analyzed the case of H$_4$, where multiple index-1 saddle points could be identified. The chemical interpretation of these points has been achieved through projection on the basis of the FCI states. In this analysis, it has been shown that some spurious index-1 saddle points arise from the introduction of nonlinearities in the linear Schrödinger equation. Only one state could be shown to correspond to an excited state, a symmetry-broken solution given by the combination of the first-excited singlet and triplet states. Therefore, some caution should be adopted when trying to assign physical meanings to the critical points of nonlinear approximation models.

A more accurate description of excited states can be achieved by post-HF methods. In particular, we aim at analyzing the CASSCF manifold through the presented theory based on Kähler manifolds. The formalization of CP and LR descriptions of CASSCF excited states will be the matter of Part II.

\section*{Acknowledgments}

This project has received funding from the European Research Council (ERC) under the European Union’s Horizon 2020 research and innovation program (grant agreement EMC2 No 810367). This work has benefited from French State support managed by ANR under the France 2030 program through the MaQui CNRS Risky and High-Impact Research program (RI)$^2$ (grant agreement ANR-24-RRII-0001). The authors are grateful to Th\'eo Duez, Simen Kvaal, Asbj\"orn Lauritsen, Tony Leli\`evre, Mathieu Lewin, Filippo Lipparini, Tommaso Nottoli, Panos Parpas, Julien Toulouse, and Jari van Gog for useful discussions.

\normalem

\printbibliography[heading=bibintoc,title={References}]

\appendix 

\section{Appendix}

\par In this section, we detail the derivations for Eqs. \eqref{eqn:FCI_excitation_energy}, \eqref{eqn:UHF CP excitation energy first order}, and \eqref{eqn:UHF LR excitation energy first order}. We also comment on the case of restricted HF. 

\subsection{Derivation for FCI first-order excitation energies}\label{subsec:FCI_derivation}

The family of parametrized energy functionals $E^{\rm FCI}(\eta,\mathbf{c})$ reads
$$\sum_{\substack{I_\sigma,J_\tau\subseteq\{1,\ldots,N_{\rm p}\}\\A_\sigma,B_\tau\subseteq\{N_{\rm p}+1,\ldots,N_{\rm b}\}\\|I_\sigma|=|A_\sigma|,~|J_\tau|=|B_\tau|,~\sigma,\tau\in\{\alpha,\beta\}}}\overline{c_{[I_\alpha,A_\alpha],[I_\beta,A_\beta]}}c_{[J_\alpha,B_\alpha],[J_\beta,B_\beta]}\left\langle\Phi_{[I_\alpha,A_\alpha],[I_\beta,A_\beta]}^{(0)}|\widehat H_\eta|\Phi_{[J_\alpha,B_\alpha],[J_\beta,B_\beta]}^{(0)}\right\rangle,$$
where $\mathbf{c}\in\mathbb{S}_{\C}^{\binom{N_{\rm b}}{N_{\rm p}}\times\binom{N_{\rm b}}{N_{\rm p}}-1}$. Here, $\Phi_{[I_\alpha,A_\alpha],[I_\beta,A_\beta]}^{(0)}$ denotes the Slater determinant formed by $(\phi_n^0\otimes\alpha)_{n\in(\{1,\ldots,N_{\rm p}\}\setminus I_\alpha)\cup A_\alpha}$ and $(\phi_n^0\otimes\beta)_{n\in(\{1,\ldots,N_{\rm p}\}\setminus I_\beta)\cup A_\beta}$; cf. Eqs. \eqref{eqn:mean field ground state} and \eqref{eqn:mean field excited states}. We denote the critical points of $E^{\rm FCI}(0,\cdot)$ by $\mathbf{c}_{[I_\alpha,A_\alpha],[I_\beta,A_\beta]}^{(0)}$ with the corresponding energies $E_{[I_\alpha,A_\alpha],[I_\beta,A_\beta]}^{(0)}$. By the choice of $(\phi_n^0)_{1\le n\le N_{\rm b}}$, $\mathbf{c}_{[I_\alpha,A_\alpha],[I_\beta,A_\beta]}^{(0)}$ is a vector with the only nonzero entry indexed by $(I_\alpha,A_\alpha,I_\beta,A_\beta)$. Note that for any $I_\alpha,I_\beta\subseteq\{1,\ldots,N_{\rm p}\}$ and $A_\alpha,A_\beta\subseteq\{N_{\rm p}+1,\ldots,N_{\rm b}\}$, $E_{[I_\alpha,A_\alpha],[I_\beta,A_\beta]}^{(0)}=E_{[I_\beta,A_\beta],[I_\alpha,A_\alpha]}^{(0)}$ due to the fact that the exchange of $\alpha$ and $\beta$ electrons does not change the energy value. As a result, each excited state is two-fold degenerate. 

\par In what follows, we focus on the first excited state. By the degenerate perturbation theory, there exist $\eta_{\rm c}>0$ and real-analytic function $(-\eta_{\rm c},\eta_{\rm c})\owns\eta\mapsto E_i^{\rm FCI}(\eta)\in\R$ and $(-\eta_{\rm c},\eta_{\rm c})\owns\eta\mapsto\mathbf{c}_i(\eta)\in\mathbb{S}_{\C}^{\binom{N_{\rm b}}{N_{\rm p}}\times\binom{N_{\rm b}}{N_{\rm p}}-1}$ such that (1) $E^{\rm FCI}_i(\eta)$ is an eigenvalue of $\widehat{H}_\eta$ with the corresponding eigenvector $\mathbf{c}_i(\eta)$ for any $\eta\in(-\eta_{\rm c},\eta_{\rm c})$ and (2) $E^{\rm FCI}_i(0)=E_{[N_{\rm p},N_{\rm p}+1],[\emptyset,\emptyset]}^{(0)}=E_{[\emptyset,\emptyset],[N_{\rm p},N_{\rm p}+1]}^{(0)}$ ($i=1,2$). Intuitively, $E^{\rm FCI}_1(\eta)$ and $E^{\rm FCI}_2(\eta)$ are two different branches starting from the same value at $\eta=0$. Expanding the FCI excitation energies in $\eta$, we obtain,
$$\omega^{\rm FCI}_i(\eta)=E^{\rm FCI}_i(\eta)-E^{\rm FCI}_{[\emptyset,\emptyset],[\emptyset,\emptyset]}(\eta)=\sum_{k=0}^{+\infty}\eta^k\omega^{{\rm FCI},(k)}_i,\quad i=1,2.$$
In particular,
$$\omega_i^{{\rm FCI},(0)}=2\sum_{j=1}^{N_{\rm p}}\varepsilon_j^0+\varepsilon_{N_{\rm p}+1}^0-\varepsilon_{N_{\rm p}}^0,\quad i=1,2.$$
To obtain $\omega_i^{\rm FCI,(1)}$, we solve the first-order FCI equations projected onto the two-dimensional subspace spanned by the vectors $\mathbf{c}_{[N_{\rm p},N_{\rm p}+1],[\emptyset,\emptyset]}^{(0)}$ and $\mathbf{c}_{[\emptyset,\emptyset],[N_{\rm p},N_{\rm p}+1]}^{(0)}$. Two distinct eigenpairs are calculated: 
\begin{itemize}
    \item the eigenvector $\frac{1}{\sqrt2}(\mathbf{c}_{[N_{\rm p},N_{\rm p}+1],[\emptyset,\emptyset]}^{(0)}+\mathbf{c}_{[\emptyset,\emptyset],[N_{\rm p},N_{\rm p}+1]}^{(0)})$ associated to the eigenvalue 
    $$\Delta\varepsilon_{N_{\rm p}+1,N_{\rm p}}^{(1)} + (N_{\rm p}+1,N_{\rm p}|N_{\rm p},N_{\rm p}+1)_{\alpha\beta};$$
    
    \item the eigenvector $\frac{1}{\sqrt2}(\mathbf{c}_{[N_{\rm p},N_{\rm p}+1,\emptyset,\emptyset]}^{(0)}-\mathbf{c}_{[\emptyset,\emptyset,N_{\rm p},N_{\rm p}+1]}^{(0)})$ associated to the eigenvalue 
    $$\Delta\varepsilon_{N_{\rm p}+1,N_{\rm p}}^{(1)} - (N_{\rm p}+1,N_{\rm p}|N_{\rm p},N_{\rm p}+1)_{\alpha\beta},$$
\end{itemize}
where $\Delta\varepsilon_{N_{\rm p}+1,N_{\rm p}}^{(1)}:=\varepsilon_{N_{\rm p}+1}^{(1)}-\varepsilon_{N_{\rm p}}^{(1)}$ and $\varepsilon_i^{(1)}$ denotes the first-order variation of the $i$-th orbital energy. One can observe from the eigenvectors that these two branches correspond respectively to the singlet excited state (the first excited state in the $\hat{S}^2=0$ sector) and a triplet excited state (the ground state in the $\hat{S}^2=3/4$ sector). We thus adopt the notations $E_{\rm S}^{\rm FCI}(\eta):=E_1^{\rm FCI}(\eta)$ and $E_{\rm T}^{\rm FCI}(\eta):=E_2^{\rm FCI}(\eta)$. Moreover, from perturbation theory and the nondegeneracy assumption, 
\begin{equation}
    \Delta \varepsilon^{(1)}_{N_{\rm p}+1,N_{\rm p}}=\braket{\Phi_{[N_{\rm p},N_{\rm p}+1],[\emptyset,\emptyset]}^{(0)}| \hat{V}|\Phi_{[N_{\rm p},N_{\rm p}+1],[\emptyset,\emptyset]}^{(0)}}-\braket{\Phi_{[\emptyset,\emptyset],[\emptyset,\emptyset]}^{(0)}| \hat{V}|\Phi_{[\emptyset,\emptyset],[\emptyset,\emptyset]}^{(0)}}
    \label{eqn:diff of first-order excitation energies}
\end{equation}
Therefore,
\begin{align*}
    \Delta\varepsilon_{N_{\rm p}+1,N_{\rm p}}^{(1)}&=\sum_{j=1}^{N_{\rm p}-1}(N_{\rm p}+1,N_{\rm p}+1||j,j)_{\beta\beta}+\sum_{j=1}^{N_{\rm p}}(N_{\rm p}+1,N_{\rm p}+1|j,j)_{\beta\alpha}\\
    &\quad-\sum_{j=1}^{N_{\rm p}}(N_{\rm p},N_{\rm p}||j,j)_{\beta\beta}-\sum_{j=1}^{N_{\rm p}}(N_{\rm p},N_{\rm p}|j,j)_{\beta\alpha}.
\end{align*}
We thus conclude that in the positive neighborhood of the origin ($\eta>0$), the first-order variation of the first excitation energy is given by the triplet excited state as follows:
\begin{align*}
\omega_{\rm T}^{\rm FCI,(1)}&=\Delta\varepsilon^{(1)}_{N_{\rm p}+1, N_{\rm p}}-(N_{\rm p}+1, N_{\rm p}|N_{\rm p},N_{\rm p}+1)_{\alpha\beta}\\
&=\sum_{j=1}^{N_{\rm p}-1}(N_{\rm p}+1,N_{\rm p}+1||j, j)_{\beta\beta}+\sum_{j=1}^{N_{\rm p}}(N_{\rm p},N_{\rm p}||j, j)_{\beta\beta} \\
&\quad-\sum_{j=1}^{N_{\rm p}}(N_{\rm p}+1,N_{\rm p}+1|j, j)_{\beta\alpha}-\sum_{j=1}^{N_{\rm p}}(N_{\rm p},N_{\rm p}|j, j)_{\beta \alpha}\\
&\quad-(N_{\rm p}+1, N_{\rm p}|N_{\rm p},N_{\rm p}+1)_{\alpha\beta}.
\end{align*}

\par We shall remark that the above arguments also apply to other excited states. The difference lies just in the two-dimensional subspace onto which the first-order FCI equations are projected. 

\subsection{Derivation for CP-UHF first-order excitation energies}

In the case of the unrestricted HF (UHF), the ground state is
\begin{equation}
\gamma_{\alpha,[\emptyset,\emptyset]}^{(0)} := \sum_{j=1}^N \phi_j^{\alpha,0} {\phi_j^{\alpha,0}}^T\qquad \gamma_{\beta,[\emptyset,\emptyset]}^{(0)} := \sum_{j=1}^N \phi_j^{\beta,0} {\phi_j^{\beta,0}}^T,
\end{equation}
and the saddle points are
\begin{align}
 &\gamma_{\alpha,[I_\alpha,A_\alpha]}^{(0)} := \gamma_{\alpha,[\emptyset,\emptyset]}^{0} - \sum_{i \in I_\alpha} \phi_i^{\alpha,0}{\phi_i^{\alpha,0}}^T + \sum_{a \in A_\alpha} \phi_a^{\alpha,0}{\phi_a^{\alpha,0}}^T \label{eqn:gamma_uhf_alpha}\\
 &\gamma_{\beta,[I_\beta,A_\beta]}^{(0)} := \gamma_{\beta,[\emptyset,\emptyset]}^{0} - \sum_{i \in I_\beta} \phi_i^{\beta,0}{\phi_i^{\beta,0}}^T + \sum_{a \in A_\beta} \phi_a^{\beta,0}{\phi_a^{\beta,0}}^T, 
 \label{eqn:gamma_uhf_beta}
\end{align}
where in each case the sets $I_\sigma \subset \{1,\cdots,N\}$, $A_\sigma \subset \{N+1,\cdots,N_{\rm b}\}$, $\#I=\#A \ge 1$ are considered to contain either orbitals with only $\alpha$ or only $\beta$ spin (it should be read as $I_\alpha, A_\alpha$ and $I_\beta, A_\beta$ for the Eqs.~\ref{eqn:gamma_uhf_alpha}-\ref{eqn:gamma_uhf_beta}, respectively). 
The first order expression for an excitation in the CP-UHF formalism is
\begin{align*}
    \omega_{[I_\alpha,A_\alpha,I_\beta,A_\beta]}^{\rm CP-UHF,(1)}&=\frac12\sum_{\kappa,\lambda,\mu,\nu=1}^{N_{\rm p}}(\kappa\lambda||\mu\nu)_{\alpha}\left(\overline{[\gamma_{\alpha,[I_\alpha,A_\alpha]}^{(0)}]_{\kappa\lambda}}[\gamma_{\alpha,[I_\alpha,A_\alpha]}^{(0)}]_{\nu\mu}-\overline{[\gamma_{\alpha,[\emptyset,\emptyset]}^{(0)}]_{\kappa\lambda}}[\gamma_{\alpha,[\emptyset,\emptyset]}^{(0)}]_{\nu\mu}\right)\\
    &+\frac12\sum_{\kappa,\lambda,\mu,\nu=1}^{N_{\rm p}}(\kappa\lambda||\mu\nu)_{\beta}\left(\overline{[\gamma_{\beta,[I_\beta,A_\beta]}^{(0)}]_{\kappa\lambda}}[\gamma_{\beta,[I_\beta,A_\beta]}^{(0)}]_{\nu\mu}-\overline{[\gamma_{\beta,[\emptyset,\emptyset]}^{(0)}]_{\kappa\lambda}}[\gamma_{\beta,[\emptyset,\emptyset]}^{(0)}]_{\nu\mu}\right)\\
    &+\sum_{\kappa,\lambda=1}^{N_{\rm p}}\sum_{\mu,\nu=1}^{N_{\rm p}}(\kappa\lambda|\mu\nu)_{\alpha\beta}\left(\overline{[\gamma_{\alpha,[I_\alpha,A_\alpha]}^{(0)}]_{\kappa\lambda}}[\gamma_{\beta,[I_\beta,A_\beta]}^{(0)}]_{\nu\mu}-\overline{[\gamma_{\alpha,[\emptyset,\emptyset]}^{(0)}]_{\kappa\lambda}}[\gamma_{\beta,[\emptyset,\emptyset]}^{(0)}]_{\nu\mu}\right)\\
    &=\frac12\sum_{\kappa,\lambda\in(\{1,\ldots,N_{\rm p}\}\setminus I_\alpha)\cup A_\alpha}(\kappa\kappa||\lambda\lambda)_\alpha-\frac12\sum_{\kappa,\lambda\in\{1,\ldots,N_{\rm p}\}}(\kappa\kappa||\lambda\lambda)_\alpha\\
    &+\frac12\sum_{\kappa,\lambda\in(\{1,\ldots,N_{\rm p}\}\setminus I_\beta)\cup A_\beta}(\kappa\kappa||\lambda\lambda)_\beta-\frac12\sum_{\kappa,\lambda\in\{1,\ldots,N_{\rm p}\}}(\kappa\kappa||\lambda\lambda)_\beta\\
    &+\sum_{\kappa\in(\{1,\ldots,N_{\rm p}\}\setminus I_\alpha)\cup A_\alpha}\sum_{\lambda\in(\{1,\ldots,N_{\rm p}\}\setminus I_\beta)\cup A_\beta}(\kappa \kappa |\lambda \lambda )_{\alpha\beta}\\
    &-\sum_{\kappa\in\{1,\ldots,N_{\rm p}\}}\sum_{\lambda\in\{1,\ldots,N_{\rm p}\}}(\kappa \kappa |\lambda \lambda )_{\alpha\beta},
\end{align*}
In particular, the lowest excitation energy is determined by the index-1 saddle points which arise from the HOMO-LUMO transition of one $\alpha$ or $\beta$ electron. Choosing (arbitrarily) the excitation to involve the $\beta$-electron, the first-order excitation energy is given by Eq.~\eqref{eqn:UHF CP excitation energy first order}.

\subsection{Derivation for LR-UHF first-order excitation energies}

Let us detail the derivation of LR first-order excitation energies using the K\"ahler manifold formalism, for the case of a system with an even number of electrons. As already mentioned above, the UHF ground state coincides with the RHF ground state for small values of $\eta$, so that LR theory involves the Hessian of the UHF energy functional at the point $x_\star(\eta):=(\gamma_\star(\eta),\gamma_\star(\eta)) \in \mathcal M={\rm Gr}(N_{\rm p},N_{\rm b}) \times {\rm Gr}(N_{\rm p},N_{\rm b})$, where $\gamma_\star(\eta)$ is the RHF $\alpha$ or $\beta$ ground-state density matrix. Recall that the RHF orbitals can be chosen real-valued since we are dealing with a non-relativistic system with no external magnetic field. It results from the real-analytic implicit function theorem that there exists $\Phi(\eta) \in  O(N_{\rm b})$ (the orthogonal group) such that 
$$
H_{\eta,\gamma_\star(\eta)} := h+2\eta J(\gamma_\star(\eta)) - \eta K(\gamma_\star(\eta)) = \Phi(\eta) \mbox{diag}(\varepsilon_1(\eta),\cdots,\varepsilon_{N_{\rm b}}(\eta)) \Phi(\eta)^T,
$$
with $\varepsilon_1(\eta) < \cdots < \varepsilon_{N_{\rm b}}(\eta)$ real-analytic, and
$$
\gamma_\star(\eta) = \Phi(\eta) \left( \begin{array}{cc} I_{N_{\rm p}} & 0 \\ 0 & 0 \end{array} \right) \Phi(\eta)^T.
$$
In addition, we have
$$
T_{x_\star(\eta)} \mathcal M = \left\{ Q=(Q^\alpha,Q^\beta), \quad Q^\sigma =   \Phi(\eta) \left( \begin{array}{cc} 0 & X^\sigma \\ {X^\sigma}^* & 0 \end{array} \right) \Phi(\eta)^T, \quad X^\sigma \in \C^{N_{\rm p} \times (N_{\rm b}-N_{\rm p})} \right\}.
$$
Using~\eqref{eq:Hessian_Grassmann} and \eqref{eqn:HF energy functional}, we see that for all $Q=(Q^\alpha,Q^\beta) \in T_{x_\star(\eta)} \mathcal M$,
\begin{align*}
{\rm Hess}_{\mathcal M} E^{\rm UHF}_\eta (x_\star(\eta)) (Q) &= {\mathcal L}_{\eta,1}(Q) + \eta {\mathcal L}_{\eta,2}(Q)
\end{align*}
with
\begin{align*}
{\mathcal L}_{\eta,1}(Q) &= \bigg( 
[ \gamma_\star(\eta),[Q^\alpha,H_{\eta,\gamma_\star(\eta)}]], [ \gamma_\star(\eta),[Q^\beta,H_{\eta,\gamma_\star(\eta)}]] \bigg), \\
{\mathcal L}_{\eta,2}(Q) &= \bigg( 
 [ \gamma_\star(\eta),[\gamma_\star(\eta),J(Q^\alpha+Q^\beta)-K(Q^\alpha)]],  [ \gamma_\star(\eta),[\gamma_\star(\eta),J(Q^\alpha+Q^\beta)-K(Q^\beta)]]
\bigg).
\end{align*}
It follows that if $Q_1=(Q_1^\alpha,Q_1^\beta),Q_2=(Q_2^\alpha,Q_2^\beta) \in T_{x_\star(\eta)}\mathcal M$, the matrix element $\langle Q_1, {\mathcal L}_{\eta,2}(Q_2) \rangle$ is given by
\begin{align*}
\langle Q_1, {\mathcal L}_{\eta,2}(Q_2) \rangle =& {\rm Re}\bigg( \int_{\R^3 \times \R^3} \frac{\rho_{Q_1}(\mathbf r) \rho_{Q_2}(\mathbf r')}{|\mathbf r-\mathbf r'|} \, d\mathbf r \, d\mathbf r' \\
& \qquad- \int_{\R^3 \times \R^3} \frac{\gamma_{Q_1}^\alpha(\mathbf r,\mathbf r') \overline{\gamma_{Q_2}^\alpha(\mathbf r,\mathbf r')}}{|\mathbf r-\mathbf r'|} \, d\mathbf r \, d\mathbf r' - \int_{\R^3 \times \R^3} \frac{\gamma_{Q_1}^\beta(\mathbf r,\mathbf r') \overline{\gamma_{Q_2}^\beta(\mathbf r,\mathbf r')}}{|\mathbf r-\mathbf r'|} \, d\mathbf r \, d\mathbf r'\bigg),
\end{align*}
where $\rho_{Q_j}$ and $\gamma_{Q_j}^\sigma$ are defined in terms of the matrices $X_j^\sigma$'s as 
\begin{align*}
& \gamma_{Q_j}^\sigma(\mathbf r,\mathbf r') := \sum_{i,a} \left( X_{ia}^\sigma \phi_i(\eta,\mathbf r) {\phi_a(\eta,\mathbf r')} + \overline{X_{ia}^\sigma} \phi_a(\eta,\mathbf r) {\phi_i(\eta,\mathbf r')} \right) \\
& \rho_{Q_j}(\mathbf r) := \gamma_{Q_j}^\alpha(\mathbf r,\mathbf r) + \gamma_{Q_j}^\beta(\mathbf r,\mathbf r).
\end{align*}
where the real-valued orbitals $\phi_n(\eta,\cdot) \in L^2(\R^3)$ are the ones constructed from the column vectors of $\Phi(\eta)$.
A suitable canonical orthonormal basis of $T_{x_\star(\eta)} \mathcal M$ is given by
\begin{align*}
\mathcal B(\eta):=& \bigg( \left(E_{[ia]}^{+\alpha}(\eta):=(\Phi(\eta) u_{[ia]}  \Phi(\eta)^* ,0) \right)_{i,a},  \left(E_{[ia]}^{+\beta}(\eta):=(0,\Phi(\eta) u_{[ia]} \Phi(\eta)^* )\right)_{i,a}, \\
& \qquad \left(E_{[ia]}^{-\alpha}(\eta):=(- \Phi(\eta) J_\star u_{[ia]} \Phi(\eta)^* ,0 )\right)_{i,a},  \left(E_{[ia]}^{-\beta}(\eta):=(0,-\Phi(\eta) J_\star u_{[ia]} \Phi(\eta)^* \right)_{i,a}) \bigg),
\end{align*}
where 
$$
u_{[ia]}:=\frac{1}{\sqrt{2}}\left( \begin{array}{cc} 0 & e_{[ia]} \\ e_{[ia]}^T & 0 \end{array} \right), ~~ J_{\star} u_{[ia]}
=\frac{1}{\sqrt{2}}\left( \begin{array}{cc} 0 & i e_{[ia]} \\ -i e_{[ia]}^T & 0 \end{array} \right), ~~ \mbox{with} ~~ [e_{[ia]}]_{jb} = \delta_{(ia),(jb)},
$$
for $1\le i,j\le N_{\rm p}$ and $N_{\rm p}+1\le a,b\le N_{\rm b}$. For these basis elements, we have
\begin{align*}
\gamma_{E_{[ia]}^{+\sigma}(\eta)}^{\sigma'}(\mathbf r, \mathbf r') &= \frac{\phi_i(\eta,\mathbf r) {\phi_a(\eta,\mathbf r')}+\phi_a(\eta,\mathbf r) {\phi_i(\eta, \mathbf r')}}{\sqrt 2} \delta_{\sigma,\sigma'}, \quad && \rho_{E_{[ia]}^{+\sigma}(\eta)}(\mathbf r) = \sqrt 2 \phi_i(\eta,\mathbf r) {\phi_a(\eta,\mathbf r)}, \\
\gamma_{E_{[ia]}^{-\sigma}(\eta)}^{\sigma'}(\mathbf r, \mathbf r') &= i\frac{\phi_i(\eta,\mathbf r) {\phi_a(\eta,\mathbf r')}-\phi_a(\eta,\mathbf r) {\phi_i(\eta, \mathbf r')}}{\sqrt 2} \delta_{\sigma,\sigma'}, \quad && \rho_{E_{[ia]}^{-\sigma}(\eta)}(\mathbf r)=0.
\end{align*}
Denoting by $\mathfrak l_{1}(\eta)$, $\mathfrak l_{2}(\eta)$, $\mathfrak h_\star(\eta)$ the matrices in this canonical basis of ${\mathcal L}_{\eta,1}$, ${\mathcal L}_{\eta,2}$ and ${\rm Hess}_{\mathcal M} E^{\rm UHF}_\eta (x_\star(\eta))$ respectively, we obtain
\begin{equation}
    \mathfrak h_\star(\eta) = \mathfrak l_{1}(\eta) + \eta \mathfrak l_{2}(\eta),
    \label{eqn:matrix rep of Riemannian Hessian UHF}
\end{equation}
with 
\begin{align*}
\mathfrak l_{1}(\eta):=\left( \begin{array}{cccc} E(\eta) & 0 & 0 & 0 \\ 0 & E(\eta) & 0 & 0  \\ 0 & 0 & E(\eta) & 0 \\ 0 & 0 & 0 & E(\eta)\end{array} \right),~~ 
\mathfrak l_{2}(\eta):= \left( \begin{array}{cccc} M_+(\eta) & N(\eta) & 0 & 0 \\ N (\eta)^T & M_+(\eta)  & 0 & 0  \\ 0 & 0 & M_-(\eta) & 0 \\ 0 & 0 & 0 & M_-(\eta) \end{array} \right),
\end{align*}
where $[E(\eta)]_{ia,jb} :=(\varepsilon_a(\eta)-\varepsilon_i(\eta)) \delta_{ij} \delta_{ab}$ and 
\begin{align*}
    [M_\pm(\eta)]_{ia,jb}=(ai||jb)_\eta\pm(ia||jb)_\eta,\quad[N(\eta)]_{ia,jb}=(ai|jb)_\eta+(ia|jb)_\eta.
\end{align*}
The subscript $\eta$ means that the two-electron integrals are calculated from the column vectors for $\Phi(\eta)$ (the RHF ground-state orbitals for $\widehat H_\eta$).
Note that in the chosen canonical basis, the symplectic matrix has the $4 \times 4$ block structure
$$
J_\star = \left( \begin{array}{cccc} 0 & 0 & I & 0 \\ 0 & 0 & 0 & I  \\ -I & 0 & 0 & 0 \\ 0 & -I & 0 & 0\end{array} \right).
$$

To proceed further, we use the following lemma.

\begin{lemma}
    Consider a family $(A(\eta))_{\eta \ge 0}$ of real symmetric $(4n) \times (4n)$ matrices of the form
    $$
A(\eta) = \left( \begin{array}{cc} A_+(\eta) & 0 \\ 0 & A_-(\eta) \end{array} \right), \quad \mbox{with} \quad A_\pm(\eta)=\Omega_0 + \eta B_\pm + \mathcal O(\eta^2), 
    $$
    and 
    $$
    \Omega_0 = {\rm diag}(\omega_1^0,\cdots \omega_n^0,\omega_1^0,\cdots,\omega_n^0) \quad \mbox{with} \quad 0 < \omega_1^0 < \cdots < \omega_n^0 .
    $$
    If the two eigenvalues $\lambda_{i,\pm}$ of the real symmetric matrix
    $$\frac12\begin{pmatrix}
        [B_++B_-]_{ii} & [B_++B_-]_{i(n+i)}\\
        [B_++B_-]_{(n+i)i} & [B_++B_-]_{(n+i)(n+i)}
    \end{pmatrix}$$
    are distinct, then for $\eta \ge 0$ small enough, the symplectic eigenvalues of $A(\eta)$ can be expanded as
    $$\omega_{i,\pm}(\eta)=\omega_i^0+\eta\lambda_{i,\pm}+\mathcal O(\eta^2),\quad i=1,\ldots,n.$$
\end{lemma}
\begin{proof}
    For $\eta \ge 0$ small enough, the symplectic eigenvalues of $A(\eta)$ are the eigenvalues of the symmetric matrix
    $$
    \Omega(\eta):=\left( A_-(\eta)^{1/2} A_+(\eta) A_-(\eta)^{1/2} \right)^{1/2}.
    $$
    Denoting by $\omega_{n+i}^0:=\omega_i^0$ for $1 \le i \le n$, we have
    \begin{align*}
    & A_-(\eta)^{1/2} = \Omega_0^{1/2} + \eta C + \mathcal O(\eta^2) \quad \mbox{with} \quad [C]_{ij}=\frac{[B_-]_{ij}}{(\omega_i^0)^{1/2}+(\omega_j^0)^{1/2}}, \\
    & A_-(\eta)^{1/2} A_+(\eta) A_-(\eta)^{1/2} = \Omega_0^2 + \eta D + \mathcal O(\eta^2), \quad \mbox{with} \quad D = \Omega_0^{3/2}C + \Omega_0^{1/2} B_+ \Omega_0^{1/2} + C \Omega_0^{3/2}, \\
    & \Omega(\eta) = \Omega_0 + \eta \Omega_1 + \mathcal O(\eta^2), \quad \mbox{with} \quad [\Omega_1]_{ij} = \frac{[D]_{ij}}{\omega_i^0+\omega_j^0}.
    \end{align*} 
    By simple algebraic manipulation, we obtain
    \begin{align*}
      [\Omega_1]_{ij} &=  \frac{1}{\omega_i^0+\omega_j^0} \left( \frac{(\omega_i^0)^{3/2} + (\omega_j^0)^{3/2}}{(\omega_i^0)^{1/2}+(\omega_j^0)^{1/2}} [B_-]_{ij} + (\omega_i^0)^{1/2} (\omega_j^0)^{1/2} [B_+]_{ij} \right) \\
      & = [B_-]_{ij} +  \frac{(\omega_i^0)^{1/2}(\omega_j^0)^{1/2}}{\omega_i^0+\omega_j^0}  [B_+-B_-]_{ij}.
    \end{align*}
   The proof is completed after applying degenerate first-order perturbation theory.
\end{proof}

Applying the above lemma to $\mathfrak h_\star(\eta)$ in Eq. \eqref{eqn:matrix rep of Riemannian Hessian UHF}, we obtain that the LR-UHF excitation energies are given by
$$
\omega_{ia,\pm}^{\rm LR-UHF}(\eta) = \varepsilon_a^0-\varepsilon_i^0 + \eta \omega_{ia,\pm}^{{\rm LR-UHF},(1)} + \mathcal O(\eta^2),
$$
where the $\omega_{ia,\pm}^{{\rm LR-UHF},(1)}$'s are obtained by diagonalizing the $2 \times 2$ matrices
$$
\left(\begin{array}{cc} \varepsilon_a^{(1)}-\varepsilon_i^{(1)} + (ai||ia)  & (ai|ia)\\ (ai|ia) &   \varepsilon_a^{(1)}-\varepsilon_i^{(1)} + (ai||ia) \end{array} \right),
$$
resulting in 
\begin{equation}
    \omega_{ia,\pm}^{\rm LR-UHF,(1)}=\varepsilon_a^{(1)}-\varepsilon_i^{(1)}+(ai||ia)\pm(ai|ia).
    \label{eqn:first-order excitation energy UHF}
\end{equation}
Note that the first-order derivative difference above is not identical to the one in the FCI level (cf. Eq. \eqref{eqn:diff of first-order excitation energies}). For any $\eta\ge0$, the Hartree-Fock equation reads
$$\left(h+\eta\big(2J(\gamma_\star(\eta)-K(\gamma_\star(\eta))\big)\right)\phi_i(\eta)=\varepsilon_i(\eta)\phi_i(\eta),\quad i=1,\ldots,N_{\rm b}.$$
For $\eta$ small enough, the $\varepsilon_i(\eta)$'s are all distinct due to the assumption. Therefore, by nondegenerate perturbation theory, 
$$\varepsilon_p^{(1)}=\left\langle\phi_p^0,(2J(\gamma_\star^0)-K(\gamma_\star^0))\phi_p^0\right\rangle=\sum_{j=1}^{N_{\rm p}}(pp||jj)+\sum_{j=1}^{N_{\rm p}}(pp|jj),\quad p=1,\ldots,N_{\rm b},$$
and
$$\varepsilon_a^{(1)}-\varepsilon_i^{(1)}=\sum_{j=1}^{N_{\rm p}}(aa||jj)+\sum_{j=1}^{N_{\rm p}}(aa|jj)-\sum_{j=1}^{N_{\rm p}}(ii||jj)-\sum_{j=1}^{N_{\rm p}}(ii|jj).$$
Plugging this back to Eq. \eqref{eqn:first-order excitation energy UHF} and using $(aa||ii)=-(ai||ia)$, we obtain
$$\omega_{ia,\pm}^{\rm LR-UHF,(1)}=\sum_{j\neq i}^{N_{\rm p}}(aa||jj)+\sum_{j=1}^{N_{\rm p}}(aa|jj)-\sum_{j=1}^{N_{\rm p}}(ii||jj)-\sum_{j=1}^{N_{\rm p}}(ii|jj)\pm(ai|ia).$$
We recover Eq. \eqref{eqn:UHF LR excitation energy first order} after letting $i=N_{\rm p}$ and $a=N_{\rm p}+1$ and taking the negative branch.
\end{document}